\documentclass[a4paper,11pt]{article}

\pdfoutput=1

\usepackage{fixltx2e}
\usepackage{jcappub} 
\usepackage{float}
  
\definecolor{RoyalBlue}{rgb}{0.25,.41,.88}

\usepackage[T1]{fontenc}
\usepackage{subfig}
\usepackage{booktabs}
\subheader{DESY 21-121}
\title{\boldmath The Effective Field Theory of Large-Scale Structure and Multi-tracer}

\author[a]{Thiago Mergulh\~ao,}
\author[b]{Henrique Rubira,}
\author[a]{Rodrigo Voivodic,}
\author[a]{L. Raul Abramo}
\affiliation[a]{\small Departamento de F\'{\i}sica Matem\'atica, Instituto de F\'{\i}sica, Universidade de S\~ao Paulo,\\ R.  do  Mat\~ao  1371,  05508-090,  S\~ao Paulo, SP, Brazil}
\affiliation[b]{\small Deutsches Elektronen-Synchrotron DESY, Notkestra\ss e 85, 22607 Hamburg, Germany}

\emailAdd{thiago.mergulhao@hotmail.com}
\abstract{We study the performance of the perturbative bias expansion when combined with the multi-tracer technique, and their impact on the extraction of cosmological parameters. 
We consider two populations of tracers of large-scale structure and perform a series of Markov chain Monte Carlo analysis for those two tracers separately. The constraints in $\omega_{\rm cdm}$ and $h$ using multi-tracer are less biased and approximately $60\%$ better than those obtained for a single tracer. The multi-tracer approach also provides stronger constraints on the bias expansion parameters, breaking degeneracies between them and with their error being typically half of the single-tracer case. Finally, we studied the impacts caused in parameter extraction when including a correlation between the stochastic field of distinct tracers. We also include a study with galaxies showing that multi-tracer still lead to substantial gains in the cosmological parameters.}


\begin{document}

\maketitle
\flushbottom

%%%%%%%%%%%%%%%%%%%%%%%%%%%%%
\section{Introduction}
\label{sec:introduction}
%%%%%%%%%%%%%%%%%%%%%%%%%%%%%

Extracting the maximum amount of information from current and forthcoming large-scale structure (LSS) surveys \cite{Abbetal,Amendola:2012ys,Benitez:2014ibt,Ivezic:2008fe} is vital to improve the bounds on cosmological parameters \cite{Beutler:2019ojk}. 
The Effective Field Theory (EFT) of large-scale structure (LSS) has been shown to be a self-consistent  framework to parametrize the theoretical uncertainties through controlled expansion parameters \cite{Baumann:2010tm, Carrasco:2012cv, Carrasco:2013mua, Konstandin:2019bay,Angulo:2015eqa}. At one-loop level, the EFTofLSS has been able to describe data with sub percent precision at (mildly) non-linear scales \cite{DAmico:2019fhj,Ivanov:2019hqk,Colas:2019ret, Philcox:2020vvt,Nishimichi:2020tvu} and is now being used to probe new physics \cite{Ivanov:2020ril,Lague:2021frh}. 

One of the main challenges relies on connecting the galaxy surveys observables with the underlying matter density field. 
For the galaxy overdensity $\delta_g$, this is done under the EFT framework by considering the relationship
\begin{equation}\label{eq:expand}
    \delta_g(\boldsymbol{x}, \tau) = \mathcal{F}[\Phi, \Phi_v]\,,
\end{equation}
where the functional $\mathcal{F}$ is expanded in terms of operators composed of derivatives of the gravitational potential $\Phi$ and of the velocity potential $\Phi_v$ \cite{Assassi2014,Desjacques2016}. As usual for EFT's, each of those operators is followed by free parameters, here referred to as bias parameters, that must be fitted when comparing the theory with data. These parameters capture both information from the tracer dynamics but also from the smallest scales (UV), with the higher-order coefficients being essential to extract information from the mildly non-linear scales. Efforts are being made to improve our knowledge about the properties of those parameters, which can lead to a more robust understanding of the aspects of tracer formation \cite{Tegmark:1998wm,Modi:2016dah,McDonald:2009dh,mirbabayi2015biased,Lazeyras:2017hxw,Schmittfull:2018yuk}. In addition to those bias coefficients, a stochastic contribution is necessary on top of Eq.~(\ref{eq:expand}) to take into account the intrinsic small-scale randomness from the initial conditions and other sources of noise  \cite{Desjacques2016}, such as sub- or super-Poissonian shot-noise, exclusion and satellite galaxies effects \cite{Baldauf:2013hka}. 

In parallel to the description of the mildly non-linear scales by the EFTofLSS, the multi-tracer (MT) technique has gathered exciting results extracting information in the linear regime. MT combines the result of different tracers to suppress the cosmic variance, maximizing the information extracted from the largest scales if compared to single-tracer (ST) \cite{Seljak:2008xr,McDonald:2008sh,Abramo2013}. MT analysis has achieved remarkable results in probing primordial non-Gaussianities and relativistic effects (in linear scales). For that reason MT has been extensively used in some recent survey analyses (see e.g., \cite{zhao2020completed,Montero-Dorta:2019kyb, favole2019cosmological}). 

In this paper we aim to combine the robustness of the EFTofLSS formalism to get deep into the non-linear regime and the gains brought by the MT analysis. 
We construct a straightforward generalization of the idea of multi-tracer within perturbation theory and perform a full-shape analysis of the power-spectrum of halos from the MultiDark N-body simulations \cite{klypin2016multidark} in real space. We leave the study of redshift-space distortions for future work. We compare the results of MT with the ST by running Markov chain Monte Carlo (MCMC) for a different set of priors in the bias parameters. Despite including many more free parameters than the ST, the MT estimate produced more stringent constraints on the cosmological and bias parameters.

This paper is structured as follows. In Sec.~\ref{sec:theory} we summarize the theoretical basis of the multi-tracer method and the large-scale bias expansion. We show how the usual ST setup, including the stochastic field and higher-derivative bias, can be extended to accommodate multiple tracers. The simulation data and the setup used to scrutinize the multi-tracer performance are given in Sec.~\ref{sec:methodology}. The main results are presented in Sec.~\ref{sec:results} and we conclude in Sec.~\ref{sec:conclusion}. We dedicate Appendix~\ref{app:taylor} to provide details of the Taylor expansion used to boost the MCMC analysis. Finally, in Appendix~\ref{app: bias_coev} we study whether co-evolution relations could be used to improve the priors. In Appendix~\ref{app:galaxies} we include an analysis with galaxies using a halo occupation distribution (HOD) to populate our halos.

%%%%%%%%%%%%%%%%%%%%%%%%%%%%%
\section{Theoretical model}
\label{sec:theory}

In this section we present the theoretical framework which underlies the multi-tracer technique and the large-scale bias expansion. 
We start by presenting the multi-tracer technique and computing the covariance of the power spectra in the Gaussian approximation. 
Next we briefly summarize the large-scale bias expansion based on the EFTofLSS, and discuss the specific details related to the bias and stochastic coefficients when considering different tracers.

\subsection{The multi-tracer technique}
\label{subsec: multi-tracer_theory}

Different tracers of the large-scale structure probe the same underlying density field, albeit in different ways. 
As a consequence, by distinguishing between tracers, we can beat down the noise introduced by cosmic variance when  we measure observables that depend on the relative clustering between the tracers \cite{Seljak:2008xr,McDonald:2008sh}. 
The idea is that, in any volume containing a realization of the Gaussian processes that give rise to the fluctuations in the density field, the statistical noise from this random process is highly correlated for all tracers of the density field, which means that we may be able to cancel cosmic variance by comparing directly the clustering of the tracers -- e.g., by taking ratios of the power spectra or correlation functions. 
Indeed, it was shown in \cite{Abramo2013} that ratios of power spectra of different tracers can be measured to an accuracy that is not constrained by cosmic variance -- only by the numbers of tracers and, of course, by the accuracy of the fiducial model. 
Moreover, in \cite{Abramo2013} it was also shown that the degrees of freedom corresponding to those ratios are completely independent (in the linear and Gaussian approximations) from the degrees of freedom corresponding to the bandpowers of the matter power spectrum -- which means that they can be measured independently.

That idea has been explored as a means to enhance the constraints on parameters from non-Gaussianities to redshift-space distortions, both in simulations and in real data \cite{2008JCAP...08..031S,2010CQGra..27l4011D,Hamaus:2010im,Hamaus:2011dq,Hamaus:2012ap,2010MNRAS.407..772G,2011MNRAS.416.3009B,ginzburg2020shot,2017PhRvD..96l3535A}.
It has also provided a blueprint for combining the clustering information from different types of galaxies in the GAMA \cite{2013MNRAS.436.3089B} and SDSS \cite{zhao2020completed,wang2020clustering,favole2019cosmological} surveys, as well as surveys on different wavelengths
\cite{2014MNRAS.442.2511F,2015ApJ...803...21B,tanidis2020developing,wang2020brief,viljoen2020constraining,liu2021coupling,gomes2020non}. 
Future surveys such as DESI \cite{2016arXiv161100036D} and J-PAS \cite{2014arXiv1403.5237B} will map large swaths of the sky using galaxies of different types (Luminous Red Galaxies, Emission Line Galaxies) as well as quasars, in order to improve constraints on cosmological parameters and to test  general relativity -- see, e.g., \cite{2020MNRAS.493.3616A}.

We now present a simplified derivation of the multi-tracer Fisher matrix, under the assumption of Gaussianity -- i.e., that the power spectrum covariance, which is a 4-point function, can be reduced to combinations of two-point functions using Wick's theorem.
For simplicity, we write the fundamental degrees of freedom as the Fourier modes of the density contrast of the tracers, together with their complex conjugates (following the notation of \cite{2016MNRAS.455.3871A})
\begin{equation}
\label{eqn:dof}
    d_i^a (\vec{k}) = \{ \tilde\delta_i (\vec{k}) \, , \, \tilde\delta_i^* (\vec{k}) \} \; ,
\end{equation}
where the index $a$ denotes each degree of freedom of $\delta$, and $i$ denotes the tracer species.
In this form, the data covariance  takes the simple form
\begin{equation}
\label{eqn:expval}
    \langle d_i^a (\vec{k}) d_j^b (\vec{k}{}') \rangle =  D^{ab}  \, C_{ij} (\vec{k},\vec{k}{}') = C^{ab}_{ij} (\vec{k},\vec{k}{}')\; ,
\end{equation}
where $D^{ab} = 1-\delta^{ab}$ and the correlation function in Fourier space is
\begin{equation}
\label{eqn:expval2}
C_{ij} (\vec{k},\vec{k}{}') 
=  \delta_{\vec{k} \, \vec{k}{}'}  \left(  P_{ij} + \delta_{ij} \, s_i \right) \; .
\end{equation}
Here $P_{ij} (\vec{k})$ is the power spectrum for tracers $i,j$, and $s_i$ is the shot noise term, which for Poisson statistics is simply $s_i = 1/\bar{n}_i$, with $\bar{n}_i$ being the number density of the species $i$.
The data covariance and the Fisher matrix are diagonal in the bandpowers as long as the Fourier space bins (the bandpowers) are sufficiently wide, $\Delta k \gtrsim \sqrt{3/2} \, V^{-1/3}$ \cite{Abramo2012, 2016MNRAS.455.3871A} for a volume $V$.
It is worth stressing that the core of the multi-tracer technique is the fact that cosmic variance is a correlated noise for all tracers, which is expressed by Eqs. (\ref{eqn:expval}) and (\ref{eqn:expval2}). 

The Fisher matrix for some set of observables $\theta^\mu$, measured in a volume $V$ and over a bandpower (Fourier bin) $\tilde{V}_k$, is given by the trace
\begin{equation}
\label{eqn:FishMat}
    F_{\mu\nu} = \frac{1}{4} V \tilde{V}_k \sum_{ijkl} \sum_{abcd} 
    \left[ C^{ab}_{ij} \right]^{-1} 
    \frac{\partial \, C^{bc}_{jk}}{\partial \theta^\mu}
    \left[ C^{cd}_{kl} \right]^{-1} 
    \frac{\partial \, C^{da}_{li}}{\partial \theta^\nu} \; ,
\end{equation}
where the extra factor of $1/2$ of the Fisher matrix is due to the fact that the Fourier modes were counted twice when we constructed our degrees of freedom with the Fourier modes and their complex conjugates. Now, using
$\left[ C^{ab}_{ij}\right]^{-1} = \left[ D^{ab}\right]^{-1} \left[ C_{ij}\right]^{-1}$, and taking into account that $\left[ D^{ab}\right]^{-1} = D^{ab}$, the second trace in the equation above gives us back the extra factor of two, and we arrive at
\begin{equation}
\label{eqn:FishMat2}
    F_{\mu\nu} = \frac{M_k}{2} \sum_{ijkl}
    \left[ C_{ij} \right]^{-1} 
    \frac{\partial \, C_{jk}}{\partial \theta^\mu}
    \left[ C_{kl} \right]^{-1} 
    \frac{\partial \, C_{li}}{\partial \theta^\nu} \; ,
\end{equation}
where $M_k = V \tilde{V}_k$ is the number of modes of the bandpower $k$ in the volume, and
\begin{equation}
 C_{ij}^{-1} = \frac{\delta_{ij}}{s_i} - \frac{1}{s_i}\frac{P_{ij}}{1+{\cal{P}}} \frac{1}{s_j} \; ,   
\end{equation}
with ${\cal{P}} = \sum_i P_{ii}/s_i$. 

We project this Fisher matrix onto the set of bandpowers of all the auto- and cross-spectra, $\{ P_{ij} \}$.
The partial derivatives in Eq. (\ref{eqn:FishMat2}) can be written as
\begin{equation}
\frac{\partial C_{ij}}{\partial P_{kl}} = \delta_{ik}\delta_{jl}
+ \delta_{il}\delta_{jk} - \delta_{ij}\delta_{jk}\delta_{kl}\delta_{li} \; ,    
\end{equation}
which results in the following expression for the Fisher matrix
\begin{equation}
\label{eqn:FishMat3}
    F_{ij,kl} \equiv F[P_{ij},P_{kl}]
     = \frac{M_k}{4} \,
    \left( 2 - \delta_{ij} \right) 
    \left( 2 - \delta_{kl} \right) \,
    \left( C_{ik}^{-1} C_{jl}^{-1} + C_{il}^{-1} C_{jk}^{-1}  \right)
    \, .
\end{equation}
Since the spectra are symmetric, $P_{ij} = P_{ji}$, this Fisher matrix has redundancies which must be eliminated, so we restrict the degrees of freedom for tracers $A$, $B$, $C$, etc., to $\theta^\mu \to \{ P_{AA},  P_{AB},  P_{AC}, \ldots, P_{BB}, P_{BC} , \ldots\} $. 
Restricting the Fisher matrix to these degrees of freedom and inverting that expression we obtain the completely general covariance matrix
\begin{align}
\label{eqn: inv_fisher}
{\cal{C}}_{ij,kl} = \frac{1}{M_k} \left( C_{ik} C_{jl} + C_{il} C_{jk} \right) \, ,
\end{align}
which results in the following terms for two tracers $A$ and $B$:
\begin{align}
\label{eqn: gaussian_cov1}
& {\cal{C}}_{AA,AA}  = \frac{1}{M_k}2P_{AA}^2N_A^2 \, ,
\\
\label{eqn: gaussian_cov3}
& {\cal{C}}_{AA,AB}  = \frac{1}{M_k}2P_{AB}P_{AA}N_{A} \, ,
\\
& {\cal{C}}_{AA,BB} = \frac{1}{M_k}2P_{AB}^2 \, ,
\label{eqn: gaussian_cov4}
\\
\label{eqn: gaussian_cov2}
& {\cal{C}}_{AB,AB} = \frac{1}{M_k}P^2_{AB} + \frac{1}{N_k}P_{AA}P_{BB}N_A N_B \, ,
\\
\label{eqn: gaussian_cov5}
& {\cal{C}}_{BB,BA}  = \frac{1}{M_k}2P_{AB}P_{BB}N_{B} \, ,
\end{align}
where $ P_{AA}$, $P_{BB}$ and $P_{AB}$ are the fiducial spectra and 
$N_{i} = 1 + s_i/P_{ii} = 1 + 1/(\bar{n}_{i} P_{ii})$ with
the assumption of Poissonian shot noise.
This multi-tracer covariance matrix will be used to combine the clustering information from two halo populations, leading to enhancements in the constraining power of that data set when compared with the case when we treat all those halos as a single tracer.

\subsection{Large-scale bias expansion}

A key step in the EFT analysis is to connect the tracers observed by galaxy surveys to the underlying dark matter density field \cite{Assassi2014}. 
One such connection is provided by Eq.~(\ref{eq:expand}), the large-scale bias expansion (see \cite{Desjacques2016} for a review), which consists in expanding the tracer density contrast over a set of operators
\begin{equation}
\label{eqn: bias_expansion}
    \delta_{g}(\boldsymbol{x}, \tau)=\sum_{\mathcal{O}} b_{\mathcal{O}}(\tau) \mathcal{O}(\boldsymbol{x}, \tau) + \epsilon(\boldsymbol{x}, \tau) + \sum_\mathcal{O}\epsilon_\mathcal{O}(\boldsymbol{x}, \tau)\mathcal{O}(\boldsymbol{x}, \tau)\,.
\end{equation}
These operators are scalars made of combinations of derivatives of the gravitational and velocity potentials ($\Phi$ and $\Phi_v$, respectively) that are invariant under Galilean transformations. In addition, an extra stochastic component $\epsilon$ and a stochastic set of operators $\epsilon_\mathcal{O}$ that are uncorrelated with the set $\mathcal{O}$ are also required. It is the case because the non-linear gravitational evolution couples the small and large modes, with the former being not accessible by the effective theory.
When connecting to observations, this set of coefficients (the bias parameters) are treated as nuisance parameters and may be fitted together with the cosmological ones \cite{Nishimichi:2020tvu}. It is important to remark that this bias expansion holds for all tracers in the sense that the set of operators is the same, but the parameters for each tracer may be different.

The most general set of operators that is relevant up to third order in the potentials and their derivatives are
\begin{equation}
\label{eqn: operator_basis}
    \mathcal{O} \in
    \left\{
    \delta, \delta^{2}, \delta^{3}, \mathcal{G}_{2}[\Phi_{g}], \delta\mathcal{G}_{2}[\Phi_{g}] , \mathcal{G}_{3}[\Phi_{g}], \Gamma_{3}[\Phi_{g}, \Phi_{v}], \nabla^2 \delta \right\}\,,
\end{equation}
where
\begin{eqnarray}
\mathcal{G}_{2}[\Phi_{\mathrm{v}}]&\equiv&[\nabla_{i j} \Phi_{\mathrm{v}}]^{2}-[\nabla^{2} \Phi_{\mathrm{v}}]^{2} \,,\\
\mathcal{G}_{3}[\Phi_{\mathrm{v}}]&\equiv&[\nabla^{2} \Phi_{\mathrm{v}}]^{3}+2 \nabla_{i j} \Phi_{\mathrm{v}} \nabla_{j k} \Phi_{\mathrm{v}} \nabla_{k i} \Phi_{\mathrm{v}}
-3[\nabla_{i j} \Phi_{\mathrm{v}}]^{2} \nabla^{2} \Phi_{\mathrm{v}}\,,\\
\Gamma_{3}[\Phi_{g}, \Phi_{v}] &\equiv& \mathcal{G}_2[\Phi_{g}] - \mathcal{G}_2[\Phi_{v}] \,.
\end{eqnarray}
We can then organize the power spectrum of a tracer in terms of the operators that contribute at each order in the loop expansion. Each n-loop term embraces the most general set of operators with n integrated momenta and their counter-terms. The power spectrum of a tracer $A$ can be written at one-loop order as \cite{Assassi2014,chudaykin2020nonlinear}:
\begin{eqnarray} 
\label{eq:generalP}
 P^{AA}(z,k) &=& \left(b^A_1\right)^2 \left[ P_0(z,k) + P_1(z,k) \right] + b^A_1b^A_2\,\mathcal{I}_{\delta^2}(z,k) + 2b^A_1b^A_{\mathcal{G}_2}\,\mathcal{I}_{\mathcal{G}_2}(z,k) \nonumber \\
 &+& \left(2b^A_1b^A_{\mathcal{G}_2} + \frac{4}{5} b^{A}_1b^A_{\Gamma _{3}}\right)\mathcal{F}_{\mathcal{G}_2}(z,k)  
+ \frac{1}{4}\left(b^A_2\right)^2\,\mathcal{I}_{\delta^2\delta^2}(z,k)  \\ &+&\left(b^A_{\mathcal{G}_2}\right)^2\,\mathcal{I}_{\mathcal{G}_2\mathcal{G}_2}(z,k) + b^A_2b^A_{\mathcal{G}_2}\,\mathcal{I}_{\delta^2\mathcal{G}_2}(z,k) + P_{\nabla^2\delta}^{AA}(z,k) + P_{\varepsilon^A\varepsilon^A}(z,k)  \,. \nonumber
\end{eqnarray}
In this expression, $P_0$ and $P_1$ are respectively the linear and one-loop contributions from Standard Perturbation Theory (SPT) \cite{Bernardeau2001} for the matter and the redshift dependence of the bias parameters was omitted for simplicity. Notice that some  operators previously presented in the basis~(\ref{eqn: operator_basis}) are eliminated from the tracer power spectrum via renormalization \cite{Assassi2014}.

We used the publicly available code \texttt{CLASS-PT} \cite{chudaykin2020nonlinear} to compute each one of the $\mathcal{I}$ and $\mathcal{F}$ terms, whose analytical expressions in terms of $P_0$ and SPT kernels can be found in Refs. \cite{Assassi2014, chudaykin2020nonlinear}. 
We also use the \texttt{CLASS-PT} built-in IR-resummation method based on \cite{Senatore:2014via}. The second-to-last term in Eq.~(\ref{eq:generalP}) comes both from the effective sound-speed of dark matter $c_s^2$ and from non-local formation of tracers at scales $R^A_*$:
\begin{equation}
    P_{\nabla^2\delta}^{AA} = -2\left[\left(b^A_1\right)^2\frac{c_s^2}{k^2_{\mathrm{NL}}} + b^A_1\left(R^A_*\right)^2\right]k^2P_{0}\,,
\end{equation}
where $k_{\rm NL}$ is the matter non-linear scale. Nevertheless, we can reabsorb $R^A_*$ and $c_s^2$ into $b^A_{\nabla^2\delta}$, so we end up with the following expression for the higher-derivative contribution
\begin{equation}
    P_{\nabla^2\delta}^{AA} = -2b^A_{\nabla^2\delta}b^A_1\frac{k^2}{k_{\rm norm}^2}P_{0}\,,
\end{equation}
where we introduced the normalizing factor $k_{\rm norm} = 0.15\, h\mathrm{Mpc}^{-1}$. In fact, we will use this same factor to normalize all $k$-dependent operators\footnote{Note that even though the choice of $k_{\rm norm}$ does not affect the results, each tracer in the multi-tracer analysis carries its own scale controlling its higher-derivative term.}. As will be described below, this redefinition is still possible when considering more tracers and cross-correlations among them. 

The last term in Eq.~(\ref{eq:generalP}) represents the stochastic piece in Eq.~(\ref{eqn: bias_expansion}) and it can be expanded as
\begin{equation} \label{eq:shotnoise}
    P_{\varepsilon^A\varepsilon^A} = \frac{1}{\Bar{n}_A}\left(1 + c^A_0 + c^A_2\frac{k^2}{k_{\rm norm}^2}\right)\,.
\end{equation}
In our analysis the Poissonian shot-noise (the first term $1/\bar{n}_A$) is subtracted from the data. In this way, the second term in the equation above embraces deviations from perfect Poissonian shot-noise \cite{Baldauf:2013hka}, whereas the third term is a scale-dependent stochastic term \cite{Eggemeier2020, Eggemeier2021}. 

\begin{figure}[h]
    \centering
    \includegraphics[width = 0.49\textwidth]{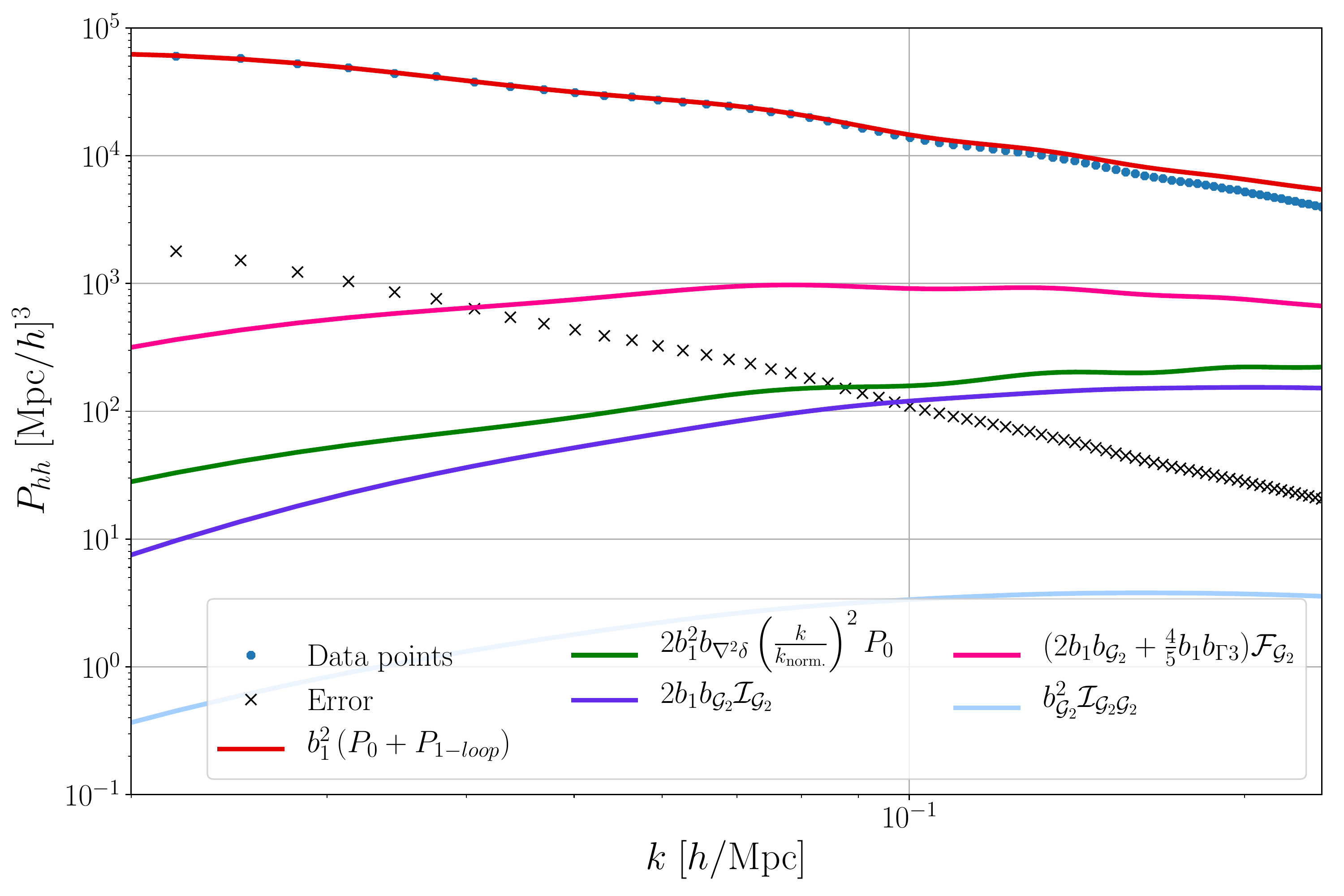}
    \includegraphics[width = 0.49\textwidth]{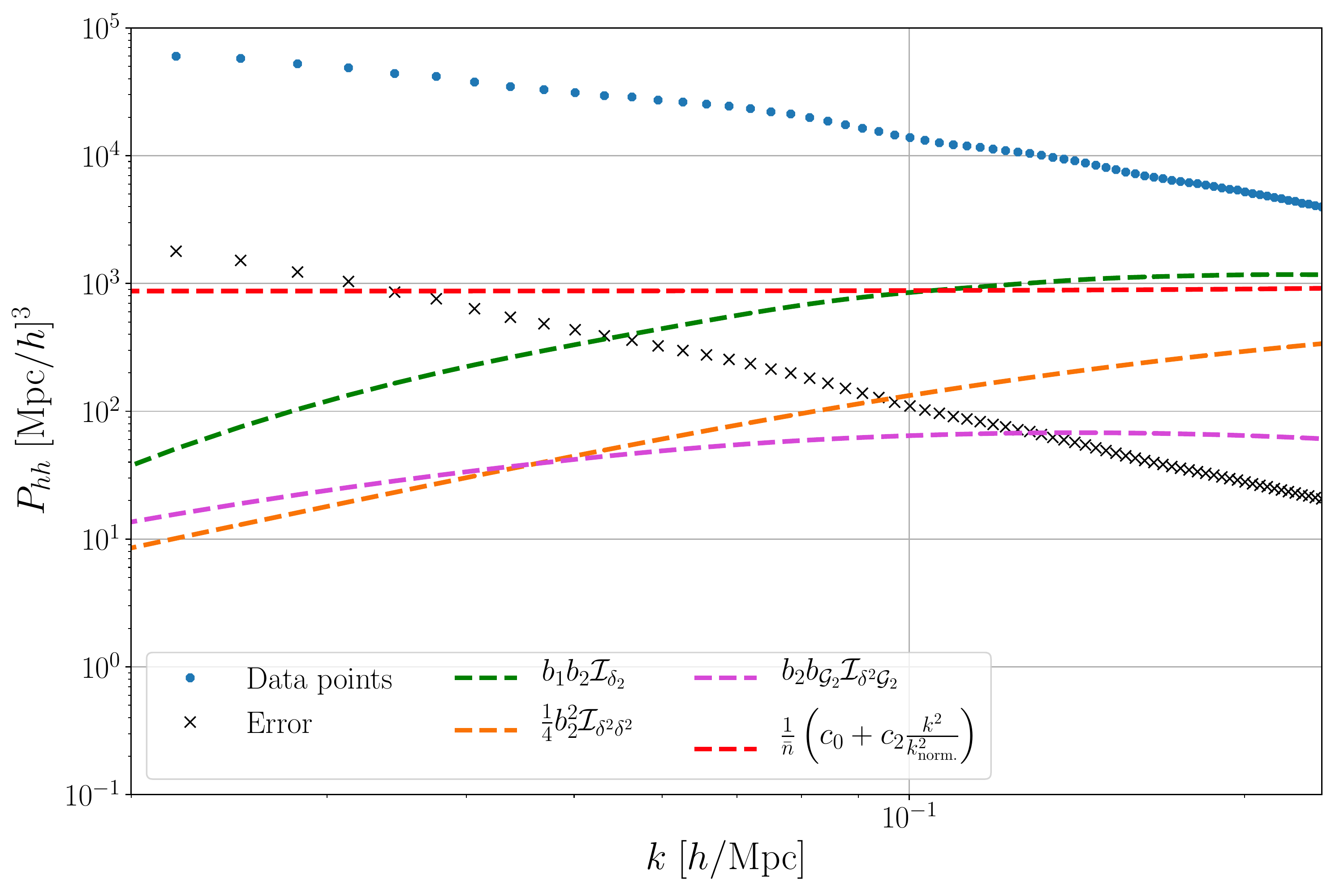}
    \caption{Each term that contributes to the one-loop power spectrum according to Eq.~(\ref{eq:generalP}). Dashed lines represent negative contributions to the power spectrum. The dots indicate the power spectrum of the halo population $A + B$ (see Table~\ref{table: halo_definition}) and the crosses indicate the estimated error from Eq.~(\ref{eqn: gaussian_cov1}). The left (right) panel shows the positive (negative) contributions.}
    \label{fig: pt_contributions}
\end{figure}

The different terms of Eq.~(\ref{eq:generalP}) are shown in Fig.~\ref{fig: pt_contributions}, where we performed a fit in the simulation data (see Table~\ref{table: halo_definition}). Since the shapes of some of these spectral corrections are very similar, some parameters are nearly degenerate.
One such degeneracy links $b_{\mathcal{G}_2}$ and $b_{\Gamma _{3}}$, such that $b_{\Gamma _{3}}$ is often neglected \cite{Ivanov2019,Wadekar:2020hax,Nishimichi:2020tvu}. 
We will keep this term in our analysis, since we want to characterize the effect of including co-evolution constraints in multi-tracer analysis (see Appendix~\ref{app: bias_coev}). Note by Fig.~\ref{fig: pt_contributions} that the most relevant contribution on the largest scales (more than the zero and one-loop terms) comes from the shot-noise\footnote{This will be relevant to explain some degeneracies found between $c_0$ and $b_1$ in Sec.~\ref{sec:results}.}, since all the other terms are built to vanish in the low $k$ regime. Notice that the overall remnant stochastic contribution to this halo population is negative. On the quasi-linear regime, the main contribution comes from  the shot-noise, $\mathcal{I}_{\delta^2}$ and $\mathcal{F}_{\mathcal{G}_2}$.

\subsubsection*{ Multi-tracer}

The simplest way to introduce multiple tracers in the model is to simply perform a bias expansions exactly as in Eq.~(\ref{eqn: bias_expansion}), providing each tracer with its own set of bias parameters. 
However, in addition to the auto-power spectra we must also model the cross-spectra, which leads us to write the bias expansion in analogy with Eq.~(\ref{eq:generalP}) as
\begin{eqnarray}
     P^{AB}(z,k) &=& b^A_1b^B_1\left[ P_0(z,k) + P_1(z,k) \right] + \frac{1}{2}\left(b^A_1b^B_2 + b^B_1b^A_2\right)\mathcal{I}_{\delta^2}(z,k)   \nonumber \\
    &+& \left(b^A_1b^B_{\mathcal{G}_2} + b^B_1b^A_{\mathcal{G}_2}\right)\mathcal{I}_{\mathcal{G}_2}(z,k)
     + \left[\left(b^A_1b^B_{\mathcal{G}_2} + b^B_1b^A_{\mathcal{G}_2}\right) 
      + \frac{2}{5}\left(b^{A}_1b^B_{\Gamma _{3}} + b^{B}_1b^A_{\Gamma _{3}}\right)\right]\mathcal{F}_{\mathcal{G}_2}(z,k) \nonumber \\ 
      &+& \frac{1}{4}b^A_2b^B_2\mathcal{I}_{\delta^2\delta^2}(z,k)  
      + b^B_{\mathcal{G}_2}b^A_{\mathcal{G}_2}\mathcal{I}_{\mathcal{G}_2\mathcal{G}_2}(z,k) + \frac{1}{2}(b^A_2b^B_{\mathcal{G}_2} + b^B_2b^A_{\mathcal{G}_2} )\mathcal{I}_{\delta^2\mathcal{G}_2}(z,k)
\nonumber \\ 
&+&P_{\nabla^2\delta}^{AB}(z,k) +  P_{\varepsilon^A\varepsilon^B}(z,k) \,,
\end{eqnarray}
The cross-term corresponding to $\nabla^2\delta$ is
\begin{eqnarray}
    P_{\nabla^2\delta}^{AB} &=& -k^2P_0
    \left[ 
    2\frac{c_s^2b^A_1b^B_1}{k_{\mathrm{NL}^2}} + b^A_1\left(R^B_{*}\right)^2 + b^B_1\left(R^A_{*}\right)^2 \right]
    \nonumber \\
    &=& \frac{1}{2}\left[\frac{b_1^B}{b_1^A}P_{\nabla^2\delta}^{AA} + \frac{b_1^A}{b_1^B}P_{\nabla^2\delta}^{BB} \right] \nonumber \\
    &=& -\left(b^A_{\nabla^2\delta}b^B_1+b^B_{\nabla^2\delta}b^A_1\right)k^2P_0 \; ,
\end{eqnarray}
such that we are able to reabsorb $c_s^2$, $R_*^A$ and $R_*^B$ into $b^A_{\nabla^2\delta}$ and $b^B_{\nabla^2\delta}$ also for the cross-spectrum. 

\begin{figure}[h]
    \centering
    \includegraphics[width = 0.85\textwidth]{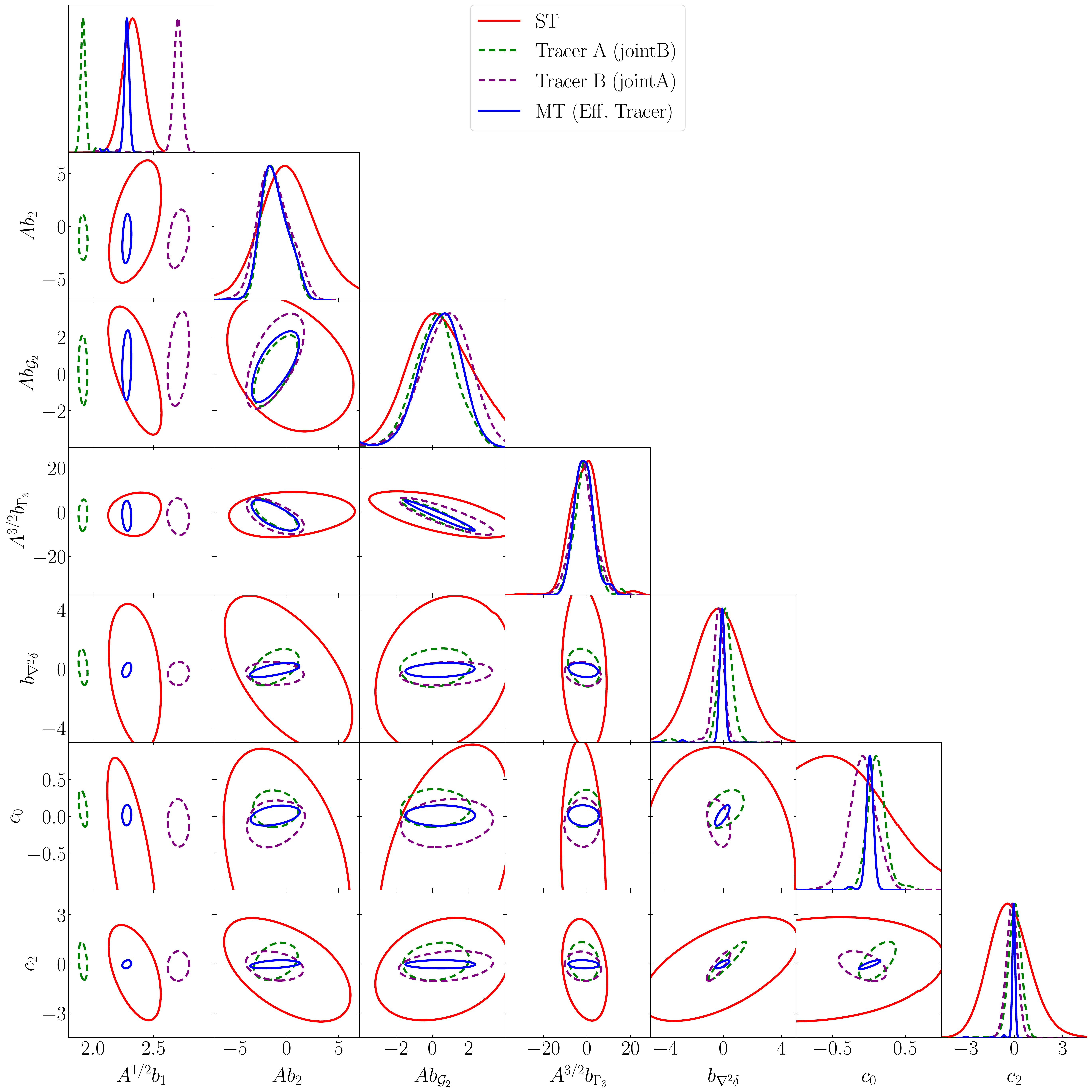}
    \caption{How the combined effective tracer parameters from multi-tracer (blue) compare to the single-tracer (red). The effective tracer bias parameters are computed using Eq.~(\ref{eqn: effective_bias}). Dashed lines indicate the values for the parameters for each tracer obtained when the fit was done together with the other tracer and their cross-spectrum. We used the $G_{0}$ priors and MCMC scheme described in Sec.~\ref{sec:methodology} and $k_{\rm max} = 0.14\, h, \mathrm{Mpc}^{-1}$.
   Notice via the definition of the effective parameters, Eqs.~(\ref{eqn: effective_bias}) and (\ref{eqn: effective_stoc}), that the combined variance for the effective tracer parameters can in principle be smaller than the variance for the individual tracers. We rescale the bias parameters by their scaling with the spectra amplitude $A$ (see Sec.~\ref{sec:paramextract}).}
    \label{fig: effective_params_example}
\end{figure}

The stochastic term of the cross-spectrum is more subtle. Tracer formation depends on the initial condition on very small scales \cite{Desjacques2016}. It is reasonable to expect that the initial conditions of two distinct tracers populations A and B are different and thus, since we are assuming Gaussian initial conditions, expect those small scales to be uncorrelated. Nevertheless, features such as the exclusion effect (the impossibility of finding the tracer A inside tracer B and vice-versa) may introduce $k^0$ and $k^2$ contributions to the noise field. Although being sub-leading when compared to the other contributions, later in this work we check its impact on the constraint of both cosmological and bias parameters. When we consider the correlation of the stochastic field of distinct tracers we will explicitly indicate. In this situation, we considered both white and non-local noise ($k^0$ and $k^2$ respectively) to the cross-correlation of the stochastic fields
\begin{equation}
    \label{eqn: cross_stoc_def}
   P_{\varepsilon^A\varepsilon^B}(z,k) = \frac{1}{\sqrt{\bar{n}_A\bar{n}_B}}\left[c^{AB}_0 + c^{AB}_2\frac{k^2}{k^2_{\mathrm{norm}}} + \mathcal{O}(k^4)\right] \, ,
\end{equation}
so they will be measured in units of the average Poissonian shot-noise of tracers $A$ and $B$.

To sum up, the two tracers ($A$ and $B$) power spectrum analysis up to one-loop, in the most complete scenario, requires a set of fourteen parameters (in addition to the cosmological parameters):
\begin{eqnarray} \label{eq:fullset}
\{b^A_1, b^A_2,b^A_{\mathcal{G}_2},b^A_{\Gamma _{3}},b^A_{\nabla^2\delta}, b^B_1, b^B_2,b^B_{\mathcal{G}_2},b^B_{\Gamma _{3}},b^B_{\nabla^2\delta}, c^{AA}_0,c^{AA}_2,c^{BB}_0,c^{BB}_2\}  \,. 
\end{eqnarray}
As mentioned above, we will also study the effect of including a stochastic term with $\{c^{AB}_0,c^{AB}_2\}$ for the cross-spectrum.

Naturally, a multi-tracer bias expansion has more bias parameters than the single-tracer case. 
To have a fair comparison between multi-tracer and a single-tracer model, it is useful to define an {\it effective tracer}  $\delta_{\rm eff}$, which is simply the union of the two species. 
The density contrast of the effective tracer is defined as
\begin{equation}
    \delta^{\rm eff} = \frac{1}{\bar{n}} 
    \sum_i \bar{n}^i\delta^i \; ,
\end{equation}
where $\bar{n} = \sum_i \bar{n}^i$ is the mean number density of the total tracer population and $\bar{n}^i$ is the number density for a specific sub-population. 
In that way, for two tracers the effective tracer parameters are given by
\begin{eqnarray}
    \label{eqn: effective_bias}
    b^{\mathrm{eff}}_{\left[\boldsymbol{\mathcal{O}}\right]} 
    &=& \frac{1}{\bar{n}} 
    \left( 
    \bar{n}^A b^{\mathrm{A}}_{\left[\boldsymbol{\mathcal{O}}\right]} + 
    \bar{n}^B b^{\mathrm{B}}_{\left[\boldsymbol{\mathcal{O}}\right]} \right) \,,
    \\
    \label{eqn: effective_stoc}
    c^{\mathrm{eff}} &=&
    \frac{1}{\bar{n}} 
    \left[
    \bar{n}^A  c^{AA} 
    + \bar{n}^B c^{BB}+ 2\sqrt{\bar{n}^A\bar{n}^B} c^{AB}
    \right] \, ,
\end{eqnarray}
where the last term in Eq.~(\ref{eqn: effective_stoc}) corresponds to the stochastic term of the cross-spectrum, which will be set to zero in most of this paper.
In Fig.~\ref{fig: effective_params_example} we show how each parameter is combined into the effective one when using the two equations above. We display the bias and stochastic parameters for each sub-species (dashed) that when combined give the MT effective tracer (blue). The ST is shown in red. Note that the effective parameters are not biased with respect to the single-tracer results. Moreover, one might find it curious that some bias parameters of individual tracers (dashed lines) might be better constrained than the full catalogue (red solid line), which in principle has more signal. Note however that the dashed lines are the values obtained by {\em jointly} fitting both sub-species A and B. We postpone a more in-depth explanation for the degeneracies and a comparison between the MT and ST schemes to Sec.~\ref{sec:results}. We now move to explain the data used and MCMC setup to compare MT to ST. 

%%%%%%%%%%%%%%%%

\section{Data and methodology}
\label{sec:methodology}

We start by presenting the simulation data and the halo catalogues used in this work. 
Next, we detail the specifications used in the MCMC analysis (priors and covariances).

\subsection{Simulation and data specifications} \label{sec:simdata}
In order to test and compare the validity of the theoretical models constructed  in Sec.~\ref{sec:theory} we perform a real space analysis using dark matter halos extracted from the MultiDark N-body simulation \cite{klypin2016multidark}. 
We use the $4\,\mathrm{Gpc}/h$ box at $z = 0$, with fiducial cosmological parameters  $\Omega_{\mathrm{m}} = 0.307$, $\Omega_{\mathrm{b}} = 0.048$, $\Omega_{\Lambda} = 0.693$, $h = 0.678$, $n_{s} = 0.96$ and $\sigma_{8} = 0.829$.

The most natural choice to construct distinct populations of tracers is to divide the halos by mass ranges. 
In this work we split the halo catalogue in two populations, $A$ and $B$, with masses below and above $13.5 \log M_{\odot}/h$, respectively.
More details on the mass ranges and number densities of those halos can be found in Table~\ref{table: halo_definition}. 
For simplicity, in this work we have limited our analysis to these two tracer species, since a sub-division into more tracers would lead to an even larger number of bias parameters.

\begin{table*}[t!h]
    \centering
      \begin{tabular}{|c||c|c|c|} \hline 
               Halo Data set & Mass range {[}$\log M_{\odot}/h${]} & $\bar{n}\,[(\textrm{Mpc}/h)^{-3}]$ & number count    \\ \hline \hline
        Halo A     & {[}13.2, 13.5{]} & $1.44 \times 10^{-4}$ & $9.21 \times 10^{6}$ \\ \hline
        Halo B     & {[}13.5, 15.7{]} & $1.23 \times 10^{-4}$ & $7.90 \times 10^{6}$ \\ \hline
        Halo A + B & {[}13.2, 15.7{]} & $2.67 \times 10^{-4}$ & $1.71 \times 10^{7}$ \\
    \hline
    \end{tabular}
    \caption{Halo populations used in the analysis. The two sub-catalogues A and B have similar number counts.}
    \label{table: halo_definition}
\end{table*}

The halo cross and auto power spectrum were measured using the usual quadratic estimator,
\begin{equation}
   P_{XY}(k_{i}) = \frac{1}{M_{k_{i}}}\sum _{k \in k_i} \tilde{\delta}_{X}(k) \, \tilde{\delta}^*
   _{Y}(k),
\end{equation}
where the bandpower $k_i$ is defined by $|k - k_{i}| < \Delta k /2$, and $M_{k_i}$ is the number of modes in that bandpower.
We employed a grid with $512$ cells per dimension (i.e., cells of $\ell=7.8125 \, h^{-1}$ Mpc), and computed $64$ bins with $\Delta k = 8\pi/L$, ranging from the fundamental mode to the Nyquist frequency ($k_0 = 2\pi/\ell$).

With the purpose of determining a fiducial value for the linear bias $b_1$ and $c_0$, we performed a simple MCMC analysis considering the following model
\begin{align}
    \label{eqn: low_k_auto}
    &P_{XX} \stackrel{k \rightarrow 0}{=} (b^X_1)^2P_{0}+ \alpha_{XX}k^2P_0 + \left(\frac{1}{\bar{n}_X}\right)c^{XX}_0\,,\\
    \label{eqn: low_k_cross}
    &P_{XY} \stackrel{k \rightarrow 0}{=} (b^X_1b^Y_1)P_{0} + \alpha_{XY}k^2P_0 +  \left(\frac{1}{\sqrt{\bar{n}_X\bar{n}_Y}}\right)c^{XY}_0 \,,
\end{align}
with some extra $\alpha$ parameters to absorb a $k^2P_0$ scaling at large scales. We fitted it up to $k = 0.12 \, h\mathrm{Mpc}^{-1}$ and their values are displayed in Table \ref{tab:fiducial_bias_shot}. 

 \begin{table}[t!h]
    \centering
      \begin{tabular}{|c||c|c|c|} \hline 
               & $b_1^X$ &             $c^{XX}_0$   & $c^{XY}_0$ \\ \hline \hline
        Halo A     & 1.288 $\pm$ 0.016  & 0.24 $\pm$ 0.15 & 0.14 $\pm$ 0.16 \\ \hline
        Halo B     & 1.806 $\pm$ 0.019    & 0.06 $\pm$ 0.21 & 0.14 $\pm$ 0.16\\ \hline
    Halo A + B & 1.528 $\pm$ 0.015     & 0.29 $\pm$ 0.31 & - \\
    \hline
    \end{tabular}
    \caption{Fiducial linear biases obtained by fitting Eqs.~(\ref{eqn: low_k_auto}) and (\ref{eqn: low_k_cross}) in the low-k regime of the galaxy power spectra.}
    \label{tab:fiducial_bias_shot} 
\end{table}

\begin{figure}[h]
    \centering
    \includegraphics[width=0.6\textwidth]{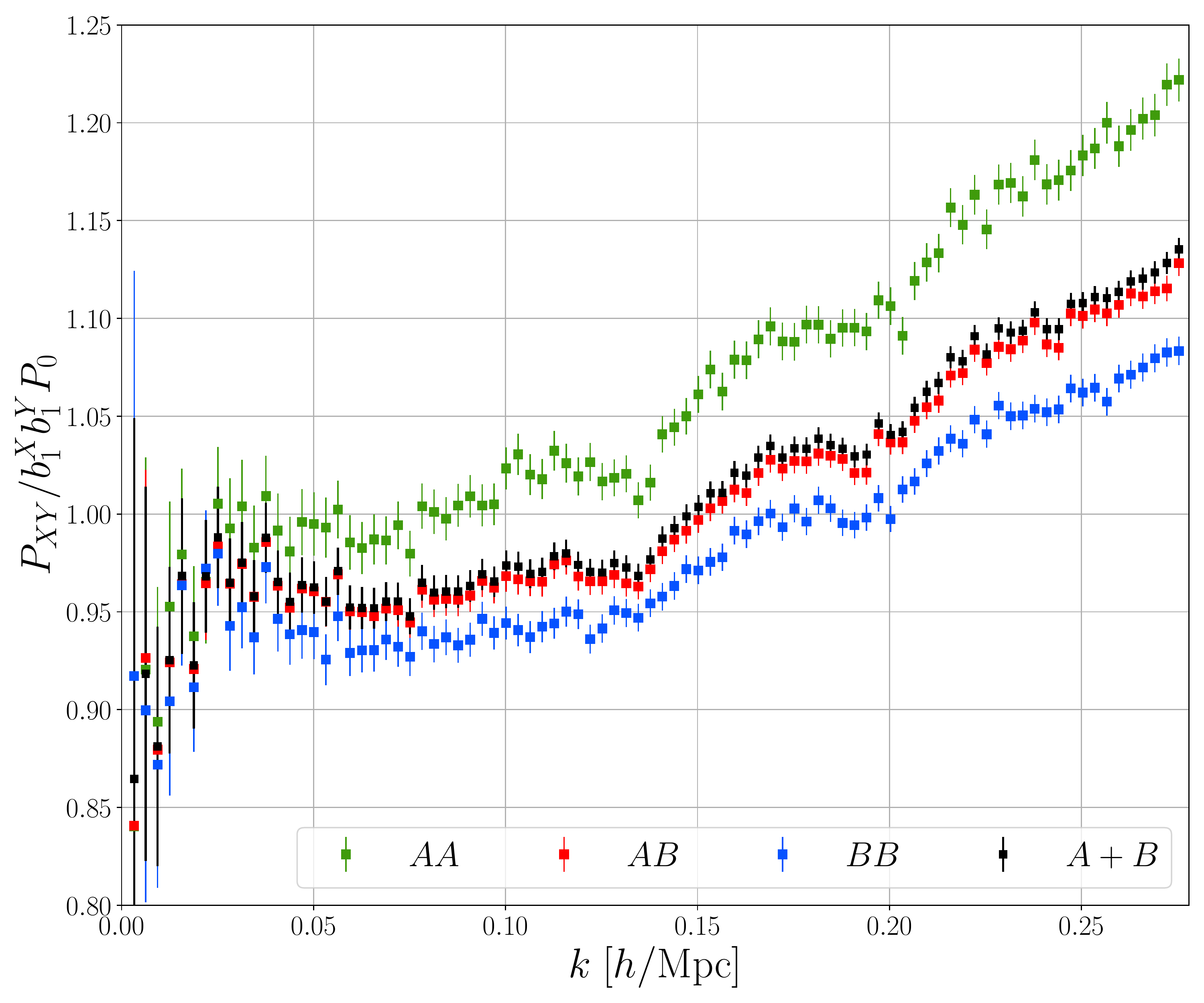}
    \caption{The power spectrum for the tracers A and B, the cross spectrum (AB) and the auto-spectrum of the full set of halos (A+B). The spectra are normalized by the linear spectrum $P_0$ and their linear biases.
    }
    \label{fig:bias_terms}
\end{figure}

In Fig.~\ref{fig:bias_terms} we show the power spectra for both tracers, normalized by the linear spectrum $P_0$ and by the respective linear bias parameters. 
It is evident that for intermediate values of $k$ the two tracers already start to develop distinct non-linear evolution. 
The different shapes for their spectrum are, in fact, the source of the additional information that can be obtained by splitting the tracers into multiple distinct populations\footnote{Note that the multi-tracer efficiency in extracting information from, e.g., redshift-space distortions or non-gaussianities, relies on the same argument: it is only when the shapes of the spectra are different that there is any information gain in their ratios \cite{Abramo2013}.}. 
We show in the same figure the cross-power spectrum ($AB$) and the auto-power spectrum of the combined halos ($A+B$). 

\subsection{Parameter extraction and data covariance matrix}  \label{sec:paramextract}

Parameter extraction is performed by standard Bayesian inference. The likelihood is assumed to be Gaussian,
\begin{equation}
\label{eqn: likelihood}
\ln\mathcal{L} 
= -\frac{1}{2}\chi^2
= -\frac{1}{2} \left(\boldsymbol{\mathcal{P}} - \boldsymbol{\mathcal{P}}_{\mathrm{data}}\right)^{t}\cdot \left[ \boldsymbol{\mathcal{C}} \right]^{-1} \cdot \left(\boldsymbol{\mathcal{P}} - \boldsymbol{\mathcal{P}}_{\mathrm{data}}\right)  \, ,
\end{equation}
where $\boldsymbol{\mathcal{P}}_{\mathrm{data}}$ and $\boldsymbol{\mathcal{P}}$ are the multi-dimensional data and theory vectors, respectively, and $\boldsymbol{\mathcal{C}}$ is the data covariance matrix.
For the multi-tracer setup we have the data vector $\boldsymbol{\mathcal{P}} = \left(\vec{P}_{AA},\vec{P}_{AB}, \vec{P}_{BB}\right)$, where arrows represent vectors of $N_k$ bandpowers, i.e.: 
$$
\vec{P}_{XX} = \left\{ P_{XX}(k_1), \cdots, P_{XX}(k_{N_k}) \right\} \; .
$$
In that way, the covariance matrix will be a $3 N_k \times 3 N_k$ square matrix constructed as
\begin{equation}
\label{eqn: 2tracer_cov}
    \boldsymbol{\mathcal{C}} = 
\begin{bmatrix} 
\mathcal{C}_{AA,AA} & \mathcal{C}_{AA,AB} & \mathcal{C}_{AA,BB} \\
\mathcal{C}_{AB,AA} & \mathcal{C}_{AB,AB} & \mathcal{C}_{AB,BB}\\
\mathcal{C}_{BB,AA} & \mathcal{C}_{BB,AB} & \mathcal{C}_{BB,BB} \\
\end{bmatrix},
\end{equation}
where $\mathcal{C}_{ij,kl}$ are computed using Eqs.~(\ref{eqn: gaussian_cov1})-(\ref{eqn: gaussian_cov4}). 
Although we have used a theoretical (Gaussian) covariance matrix, a more accurate computation using simulations typically does not impact significantly the constraints on cosmological parameters -- see, e.g., \cite{Blot_2019}. The diagonal part of the covariance matrix is typically dominated by two contributions: the shot-noise and the Gaussian terms. MT typically increases the shot-noise term of the auto-correlation which is already embraced by our theory. In the cross-correlation, the absence of a shot-noise term can make the non-Gaussian contribution more relevant. Though, the Gaussian part is still the most relevant, being five times larger in the regime $k_{\rm max} < 0.2 h\mathrm{Mpc}^{-1}$ considered in this work (see \cite{Wadekar:2019rdu}). The non-Gaussian term contributes more for the non-diagonal part of the covariance but it does not degrade the constrain power of the ST \cite{Blot_2019} (in the mild non-linear regime) and we do not have any \it a priori \rm reason to believe it would be different in the MT scenario. Finally, in a real survey there will be a series of corrections to the covariance matrix, e.g. from the survey mask and the selection function. So ultimately a proper covariance should be computed on the basis of simulations of the real survey.

\begin{table*}[t!h]
    \centering
    %%%%%%%%%%%%%%%%%%%%%%%%%%%%%%%%%%%%%%%%%%
          \begin{tabular}{|c||c|} \hline 
        &  Prior    \\ \hline \hline
        $\omega_{cdm}$  & Flat $[0.095, 0.14]$\\ \hline
        $h$   & Flat $[0.6, 0.75]$ \\ \hline
    $A$   & Flat $[1.49, 2.8]$\\
    \hline
    \end{tabular}
    \caption{Priors over the cosmological parameters.}
    \label{table:priors_cosmo}
\end{table*}

\begin{table*}[t!h]
    \centering
    %%%%%%%%%%%%%%%%%%%%%%%%%%%%%%%%%%%%%%%%%%
    \begin{tabular}{|c||c|c|} \hline 
     &  Prior \textit{Flat} &  Prior \; $G_0(\sigma)$ \\ \hline \hline
    $b_1$   & \multicolumn{2}{|c|}{Flat $[1.0, 2.2]$}    \\ \hline
    $b_2$   &  Flat $[-5.0, 5.0]$ & Gauss.(0,$\sigma$)  \\ \hline 
    $b_{\mathcal{G}_2}$   &  Flat $[-5.0, 5.0]$ &  Gauss(0,$\sigma$)  \\ \hline 
    $b_{\Gamma_3}$   &  Flat $[-10.0, 10.0]$ &  Gauss(0,2$\sigma$)  \\ \hline 
    $b_{\nabla^2\delta}$   &  \multicolumn{2}{|c|}{Flat $[-5.0, 5.0]$}  \\ \hline
    $c_0$   &  \multicolumn{2}{|c|}{\,\,\,\,Flat $[-5.0, 5.0]$}  \\ \hline 
    $c_2$   &  \multicolumn{2}{|c|}{Flat $[-5.0, 5.0]$}  \\ \hline 
    \end{tabular}
    \caption{Priors over the bias and stochastic parameters for the three scenarions considered in our analysis: \it Flat \rm and $G_0(\sigma)$.}
    \label{table:all_priors}
\end{table*}

The posterior was sampled using the \texttt{emcee} \cite{2013PASP..125..306F} Python package  with the standard move proposal\footnote{This move proposal has a free parameter, the \textit{stretch scale}, denoted by $a$ in the \texttt{emcee} documentation. The default value is $a = 2.0$. However, during our analysis, we concluded that $a$ must be changed to smaller values in order to obtain a satisfactory acceptance ratio. For the single-tracer case the typical values were $a = 1.5-1.7$, whereas for multi-tracer we used $a = 1.3-1.5$.}. The convergence was monitored using the Gelman-Rubin criteria. For the scale reduction factor we adopted the value $\epsilon = 0.03$. The chain analysis was performed using the \texttt{getdist} library \cite{Lewis:2019xzd}.

An essential part of Bayesian parameter inference is the choice of priors.
In this work, we consider the cosmological parameters $\{ \omega_{\rm cdm}, h, A_s\}$\footnote{This specific choice of parameters is in tandem with the constrain power of the full-shape analysis. As described later in the text, the main gains of MT come from the break of degeneracies in the spectra shape and we here focus on isolating this effect. A more-in-depth discussion of other effects, such as Baryon Acoustic Oscillation and  Alcock-Paczynski is left for a future work.} under the flat priors detailed in Table~\ref{table:priors_cosmo}. We rescale the primordial amplitude as $A_s \equiv A\times 10^{-9}$. Even though we let $A$ vary in the analysis, as we restrained this work to the real space, the constraint power over $A$ is minimal and we focus on $\omega_{\rm cdm}$ and $h$. All the others parameters are fixed to their fiducial values used in the MultiDark simulations. For each set of cosmological parameters we compute $P_0$, $P_1$ and the $\mathcal{I}$ and $\mathcal{F}$ terms from Eq.~(\ref{eq:generalP}) using the IR-ressummed version of \texttt{CLASS-PT} \cite{chudaykin2020nonlinear}.
We follow \cite{Colas2019} and boost the parameter space exploration by Taylor expanding the calculation of the loop corrections, as described in Appendix \ref{app:taylor}.

Many distinct prior choices for the bias and stochastic parameters have been previously considered in the literature. In this paper, we address a more careful study of the effect of the prior choice and how it is affected by multi-tracer analysis. Our main results are obtained considering two distinct prior setups, as in Table~\ref{table:all_priors}. The first one is a conservative flat prior over all parameters, labeled as \textit{Flat}. The second one considers Gaussian priors centered at zero for $b_2, b_{\mathcal{G}_2}$ and $b_{\Gamma_3}$. We label this prior structure as $G_0(\sigma)$. For $b_{\Gamma_3}$ we consider a variance that is twice the variance of the other parameters, since the typical range of variation of this parameter is higher\footnote{That can be understood in terms of perturbation theory: $b_{\Gamma_3}$ is a higher-order term whose contribution might be more relevant on more non-linear scales. It is therefore less constrained than the other terms.}. We take as fiducial value $\sigma = 2$ and leave a more in-depth study of the effect of changing $\sigma$ to Appendix~\ref{app: bias_coev}. Note that in all the scenarios we use a flat prior for $b_1$, $b_{\nabla^2 \delta}$, $c_0$ and $c_2$. The very conservative boundaries chosen in that case certificate that the typical expected values for those parameters are quite within those limits. We also consider a prior using the bias co-evolution relations, which are theoretical relations of the higher-order bias $b_{\mathcal{G}_2}$ and $b_{\Gamma_3}$ with $b_1$ (see Eqs. \ref{eqn: bias_coev_bg2} and \ref{eqn: bias_coev_bGamma3}). However, to avoid clutter the discussion, we move these results to Appendix \ref{app: bias_coev}. 

When varying the spectral amplitude it is necessary to rescale the bias parameters in order to reduce the degeneracies between the parameters \cite{Eggemeier2021}. Therefore, we make the change
\begin{equation}
    b_{[\boldsymbol{\mathcal{O}_N}]} \rightarrow b_{[\boldsymbol{\mathcal{O}_N}]}A^{N/2}\,,
\end{equation}
where $N$ denotes the order of that specific operator in the bias expansion. We applied this to $b_1, b_2, b_{\mathcal{G}_2}$ and $b_{\Gamma_3}$.

\section{Results}
\label{sec:results}

Now we compare the constraining power between the single-tracer case, with seven bias and stochastic parameters, and the multi-tracer, with the fourteen parameters described by~(\ref{eq:fullset}). 
We perform a full-shape power-spectrum analysis through a set of MCMC runs, according to Sec.~\ref{sec:methodology}. We study how the constraining power of the MT analysis is compared to ST for different values of $k_{\rm max}$, the maximum value of $k$ for which we fit the data, corresponding to the smallest scales that we assume the EFT setup to provide reliable and unbiased results. 
By increasing $k_{\rm max}$ we tend to shrink the error bars in the parameters, but pushing the theory to very small scales may also bias the parameters. 

\begin{figure}[h]
    \centering
    \includegraphics[width = 0.85\textwidth]{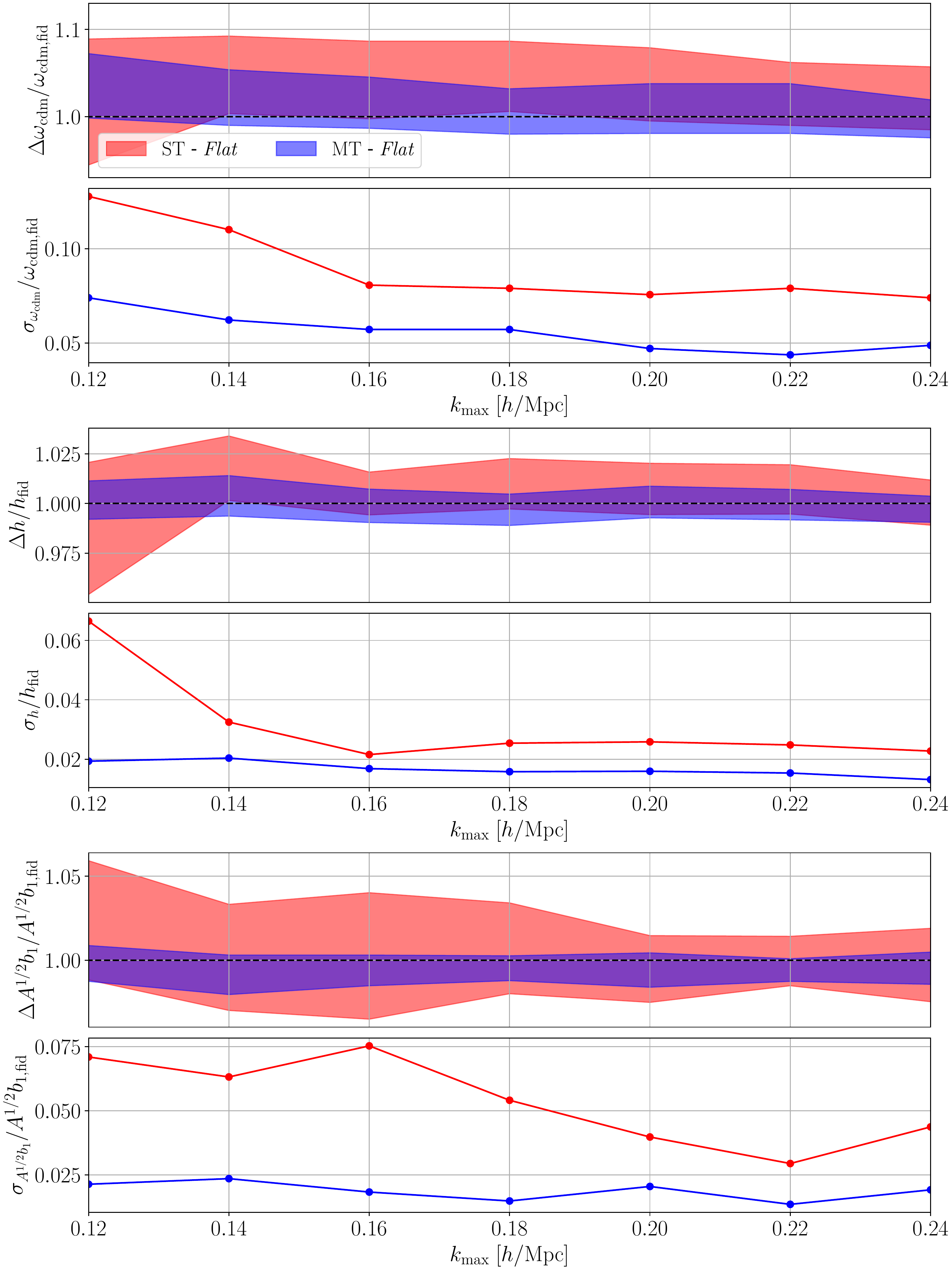}
    \caption{The error in $\omega_{\rm cdm},\,h$ and  $A^{1/2}b_1$ as a function of $k_{\rm max}$ using the \textit{Flat} prior setup for single-tracer (red) and multi-tracer (blue). The color bands indicate the $1\sigma$ confidence region. The line plot indicates the error bars normalized by the fiducial value of the parameter.}
    \label{fig: flat_result}
\end{figure}

The priors we considered are described in Table \ref{table:all_priors}, both for MT and ST. In Fig.~\ref{fig: flat_result} we show the 1$\sigma$ intervals for $\omega_{\rm cdm},\, h$ and $b_1$ obtained when using the \textit{Flat} prior. 
As previously mentioned, in real space the constraining power for $A$ is very limited by using only clustering information, and for this reason we will not consider it in our results. For most of the range of $k_{\rm max}$ considered here, we find that the MT produced unbiased results. The value of $\omega_{\rm cdm}$ gets a slight level of bias for ST. As expected, the relative sizes of the error bars shrink as a function of $k_{\rm max}$ for both MT and ST. However, the MT error bars for the cosmological parameters are on average $60\%$ smaller than those obtained with the ST analysis. 
The most striking improvement occurs when considering only the largest scales: for $k \lesssim 0.16 \, h$Mpc$^{-1}$ the MT improvements in the \textit{Flat} scenario are up to $\sim 80\%$ for $\omega_{\rm cdm}$ and $\sim 200\%$ for $h$ and $b_1$.

Note however that combining flat priors for two tracers does not exactly lead to a flat prior on the effective tracer. This can be understood from the fact that the sum of two uniformly distributed variables does not lead to a uniformly distributed random variable. In that sense, one could argue that we would be using the prior twice in the MT case if we compared it to the ST case. 
We can scrutinize this hypothesis by using Gaussian priors. From Eq.~(\ref{eqn: effective_bias}) we see that a Gaussian prior on the individual tracers can be mapped onto a Gaussian prior for the combined tracer as:
\begin{align}
    \label{eqn: compensating_factor}
    (\bar{n}_A+\bar{n}_B)^2 \sigma_{b,A+B}^2 = 
    \bar{n}_A^2 \sigma_{b,A}^2  + \bar{n}_B^2 \sigma_{b,B}^2 .
\end{align}
Therefore, if the individual tracers have approximately equal densities, a prior $\sigma_{b,A} = \sigma_{b,B}$ would be equivalent to a prior $\sigma_{b,A+B} \simeq 0.71 \, \sigma_{b,A}$. With this in mind, we compensated the ST $\sigma$ in the Gaussian prior by that extra factor.

\begin{figure}[h]
    \centering
    \includegraphics[width = 0.85\textwidth]{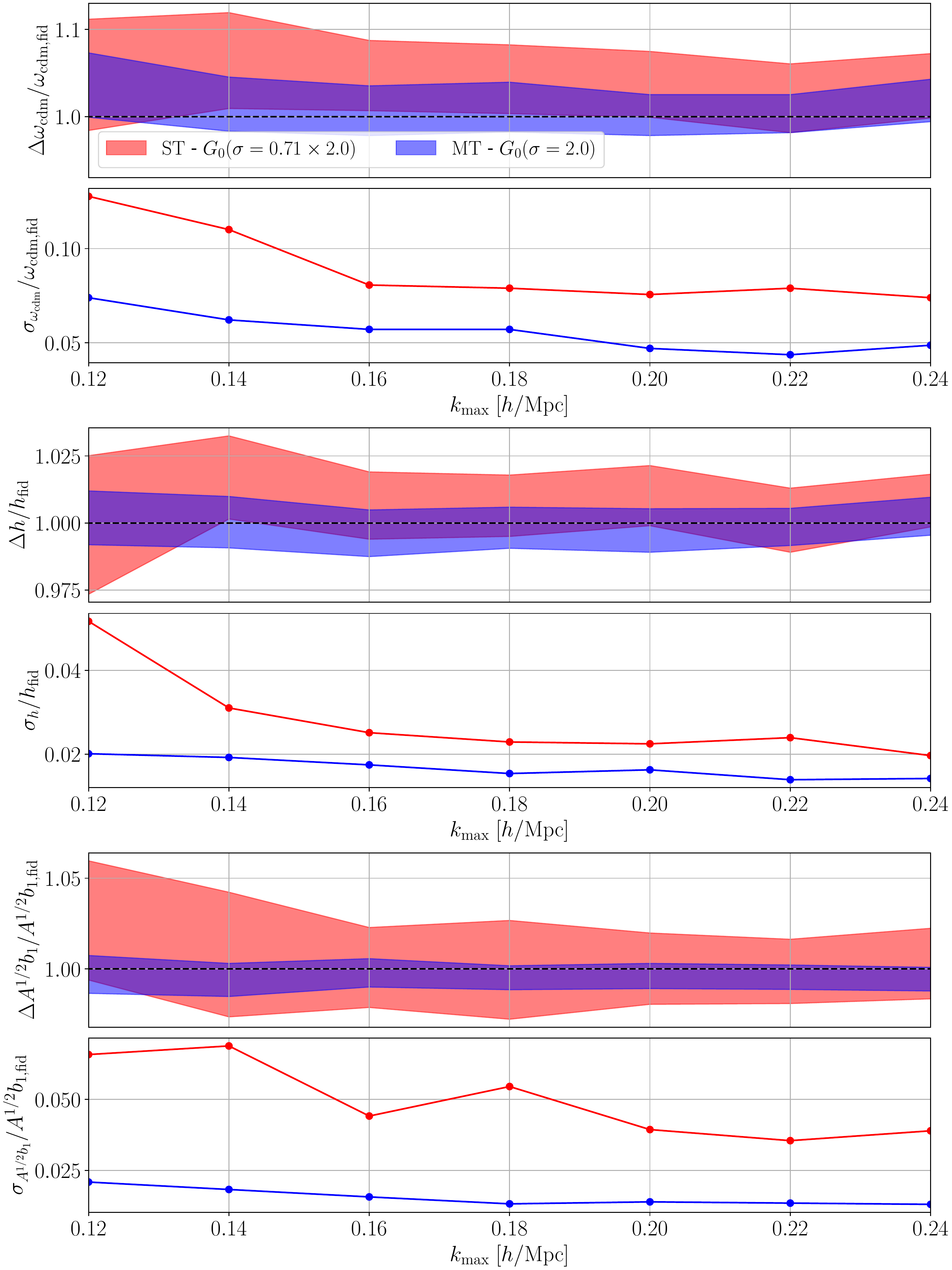}
    \caption{Same as Fig.~\ref{fig: flat_result}, but for the $G_{0}(\sigma)$ prior. We used the compensating factor for $\sigma$ in the ST prior (see discussion around Eq.~(\ref{eqn: compensating_factor})).}
    \label{fig:G0prior_results}
\end{figure}

In Fig.~\ref{fig:G0prior_results} we present the results using the Gaussian priors $G_0$ (see Table~\ref{table:all_priors})\footnote{Note that for some parameters we still use a flat prior with a broad range to guarantee that they are not affected by border effects in the convolution of the two tracers.}.  We included the compensating factor for ST and fixed $\sigma = 2$ which has shown to be a value for which the error bars are stable. For a more in-depth discussion about the effects of $\sigma$, see Appendix~\ref{app: bias_coev}. 
Note that for the considered range of $k_{\rm max}$, we again did \it not \rm find any significant biasing in the cosmological parameters and $b_1$ for MT. Still, $\omega_{\rm cdm}$ is slightly biased in the ST analysis for intermediate $k_{\rm max}$. The estimated error bars for MT are, however, consistently narrower than for ST: for $\omega_{\rm cdm}$ and $h$ the MT error bars are approximately $60\%$ of those obtained using ST. For $b_1$ the improvement is even better, and the reason for this will be made clear below. Therefore, even with the compensating factor, the MT analysis still provides better constraints than ST, and the results for $G_0$ assimilate to those obtained using the \textit{Flat} priors. 

\begin{figure}[h]
    \centering
    \includegraphics[width = \textwidth]{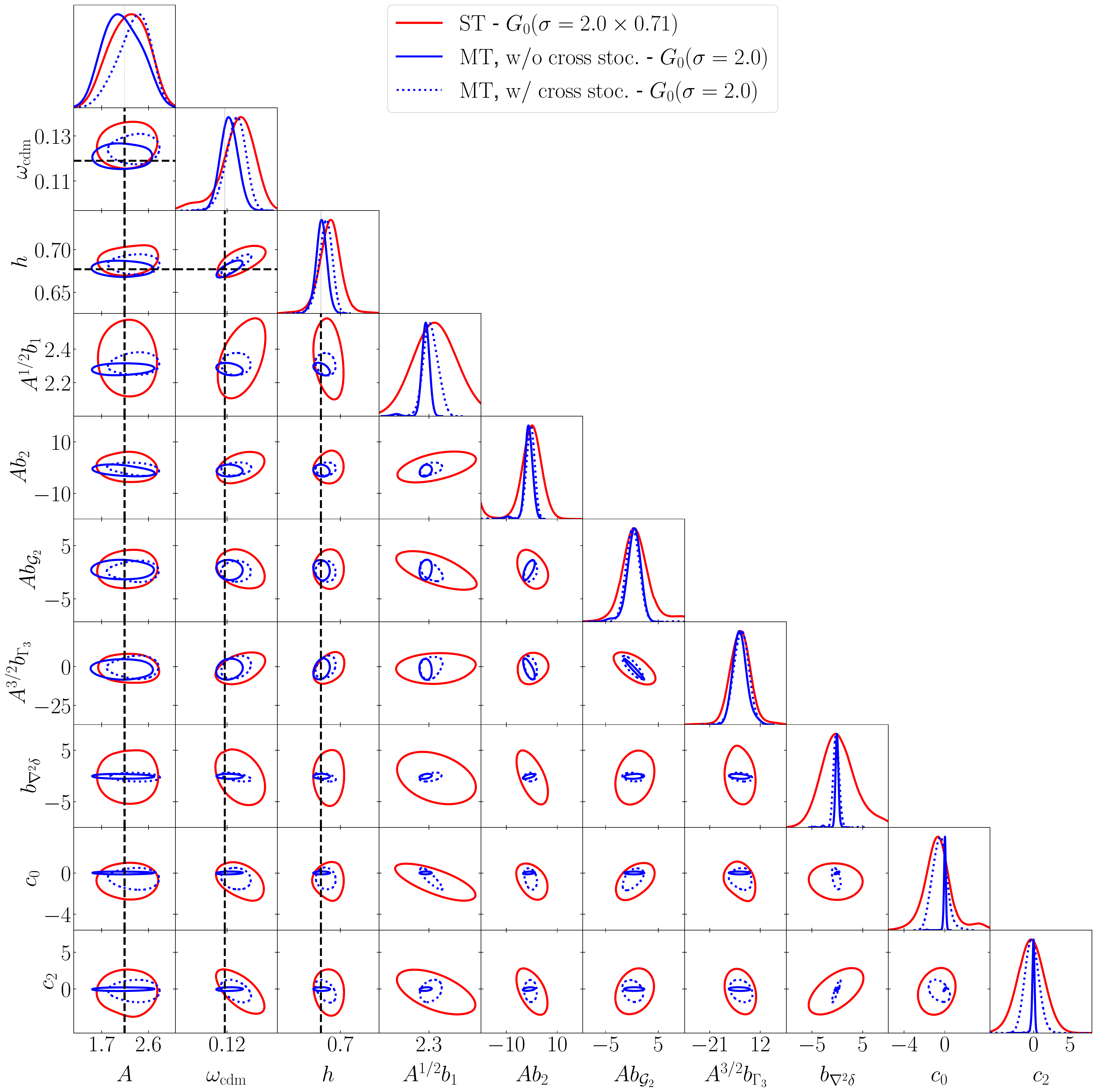}\caption{The $68\%$ confidence intervals of the sampled posterior for $G_{0}(\sigma)$ with and without the multi-tracer technique. We fixed $k_{\rm max} = 0.14  \,h\mathrm{Mpc}^{-1}$. For the MT result we are showing the effective bias and stochastic parameters, defined in Eqs.~(\ref{eqn: effective_bias}) and (\ref{eqn: effective_stoc}), respectively.}
    \label{fig:mcmc_result_G0}
\end{figure}

In order to study the constraining power of MT while considering the full set of parameters, we show in Fig.~\ref{fig:mcmc_result_G0} the combined posteriors obtained from the  MCMC for $k_{\rm max} = 0.14 \, h\mathrm{Mpc}^{-1}$ using the $G_0$ priors. The two-loop contribution starts to become more relevant on scales  $k_{\mathrm{max}} > 0.14\, h\mathrm{Mpc}^{-1}$ \cite{Baldauf:2016sjb,Nishimichi:2020tvu}, hence we adopt this value as our fiducial $k_{\mathrm{max}}$ for the following figures.\footnote{Other works as \cite{Nishimichi:2020tvu} could see a deviation from the fiducial parameters for large $k_{\rm max}$. It was not observed for our results probably due to the relatively smaller box size of MultiDark simulation.} We compare the ST (red) with the MT (blue), and we also consider a scenario including a stochastic term in the cross-spectrum for MT (blue dashed) with $c_0^{AB}$ and $c_2^{AB}$ -- for the priors, look at Table~\ref{table:all_priors}. We can see substantial improvements in the bias and stochastic parameters when considering MT, both with and without the stochastic terms. Nevertheless, the addition of $c^{AB}_0$ and $c^{AB}_2$ increases the uncertainties of some parameters. In the presence of a stochastic term in the cross-spectrum, the constraining power of MT is slightly reduced, especially for some of the bias parameters. The exclusion of a cross stochastic term, however, is very well justified, since we expect the small-scale stochastic effects of the two tracers to be uncorrelated\footnote{As previously mentioned, some non-leading order effects might still contribute, such as the exclusion effect for the two halo species.}. When measuring $c_0^{AB}$ using Eq.~(\ref{eqn: low_k_cross}) (see Table~\ref{tab:fiducial_bias_shot}) we find a value for the cross stochastic term that is $1-\sigma$ consistent with zero. This, when added to the fact that the halo model \cite{Cooray:2002dia} predicts a value that is considerably smaller than all other stochastic terms \cite{Hamaus:2010im}, lead us to neglect this term in our main results. We studied the parameter space and concluded that including the cross-stochastic term as free parameter can mildly impact the multi-tracer performance (see Figure \ref{fig:mcmc_result_G0}), indicating that one should look for well-motivated priors in order to better leverage the available information, akin done by \cite{2021arXiv211200012K}. We include in Appendix~\ref{app:galaxies} an analysis with galaxies that include the cross-stochastic term, in which we populate our halos using a halo occupation distribution. Our results are still similar in that case.

Note from Fig.~\ref{fig:mcmc_result_G0} that the posteriors for $b_1$, $b_{\nabla^2 \delta}$, $c_0$ and $c_2$ become drastically narrower using MT compared with the ST scenario, especially when we neglect the stochastic terms. 
The reason for that is the following: in the ST scenario the linear bias $b_1$ and the stochastic term $c_0$ present a high degree of degeneracy. 
This can be understood by inspecting Fig.~\ref{fig: pt_contributions}, and noticing that at very low $k$ the dominant terms are in fact the linear power spectrum and the stochastic term. Therefore, a positive shift in $b_1$ can be absorbed by a negative shift in $c_0$ or vice-versa, with the other bias parameters  compensating those shifts at intermediate scales.

\begin{figure}[h]
    \centering
    \includegraphics[width = 0.8\textwidth]{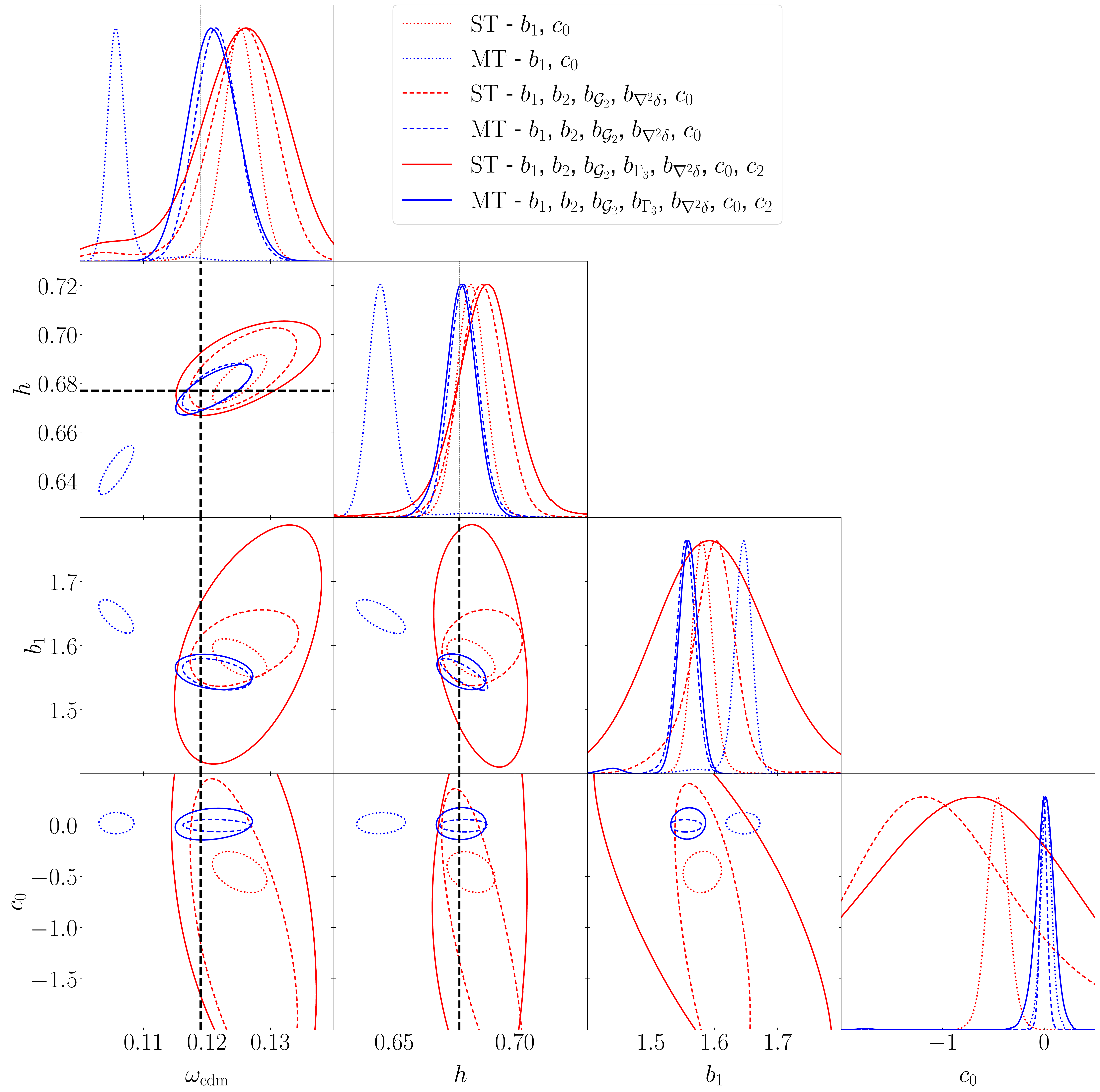}
    \caption{The $68\%$ confidence curves for $k_{\mathrm{max}} = 0.14\, h\mathrm{Mpc}^{-1}$ for three distinct models (dotted, dashed and solid lines) for ST (red) and MT (blue). We used the $G_{0}(\sigma)$ prior, based in Table \ref{table:all_priors}, and the compensating factor for the single-tracer as discussed in Eq.~(\ref{eqn: compensating_factor}).}
    \label{fig:distinct models}
\end{figure}

We display in Fig.~\ref{fig:distinct models} the combined posterior for $h$, $\omega_{\rm cdm}$, $b_1$ and $c_0$ for three different models fitted until $k_{\mathrm{max}} = 0.14\, h\mathrm{Mpc}^{-1}$ using the $G_0$ prior. 
The dotted lines denote fits to a linear model including only $b_1$ and $c_0$, for both MT (blue) and ST (red). We progressively add more free parameters in the models described by dashed and solid lines, the latter being the full model considered in our analysis. We can see that both for the minimal linear case (dotted lines) and the model described by dashed lines, there is almost no degeneracy between $c_0$ and $b_1$ for ST. The inclusion of $\Gamma_3$ and, in particular, $c_2$ give enough freedom to the system to start developing the degeneracy between $c_0$ and $b_1$. 

Moreover, we notice from Fig.~\ref{fig:distinct models} that the benefits from using MT are significantly lower when considering only the simplest model (linear fit, dotted lines). 
As we add more freedom to the system, we of course expect the constraints on physical parameters to relax, at the same time that they become less biased. This deterioration in the MT scenario, however, is less present than in the ST. The constraining power of MT, in that sense, is more powerful: it captures extra information out of non-linear parameters, these parameters become less degenerate, and the cosmological parameters get less biased when compared to the ST analysis. 
Note that if we consider the simplified scenario without $b_{\Gamma_3}$ and $c_2$, very often considered in the literature and represented in Fig.~\ref{fig:distinct models} by dashed lines, the ST produces relatively smaller error bars and unbiased results. In that scenario, the MT beats the ST by a factor of $43 \%$ for $\omega_{\rm cdm}$ and $42\%$ for $h$.

\begin{figure}[h]
    \centering
    \includegraphics[width = 0.49 \textwidth]{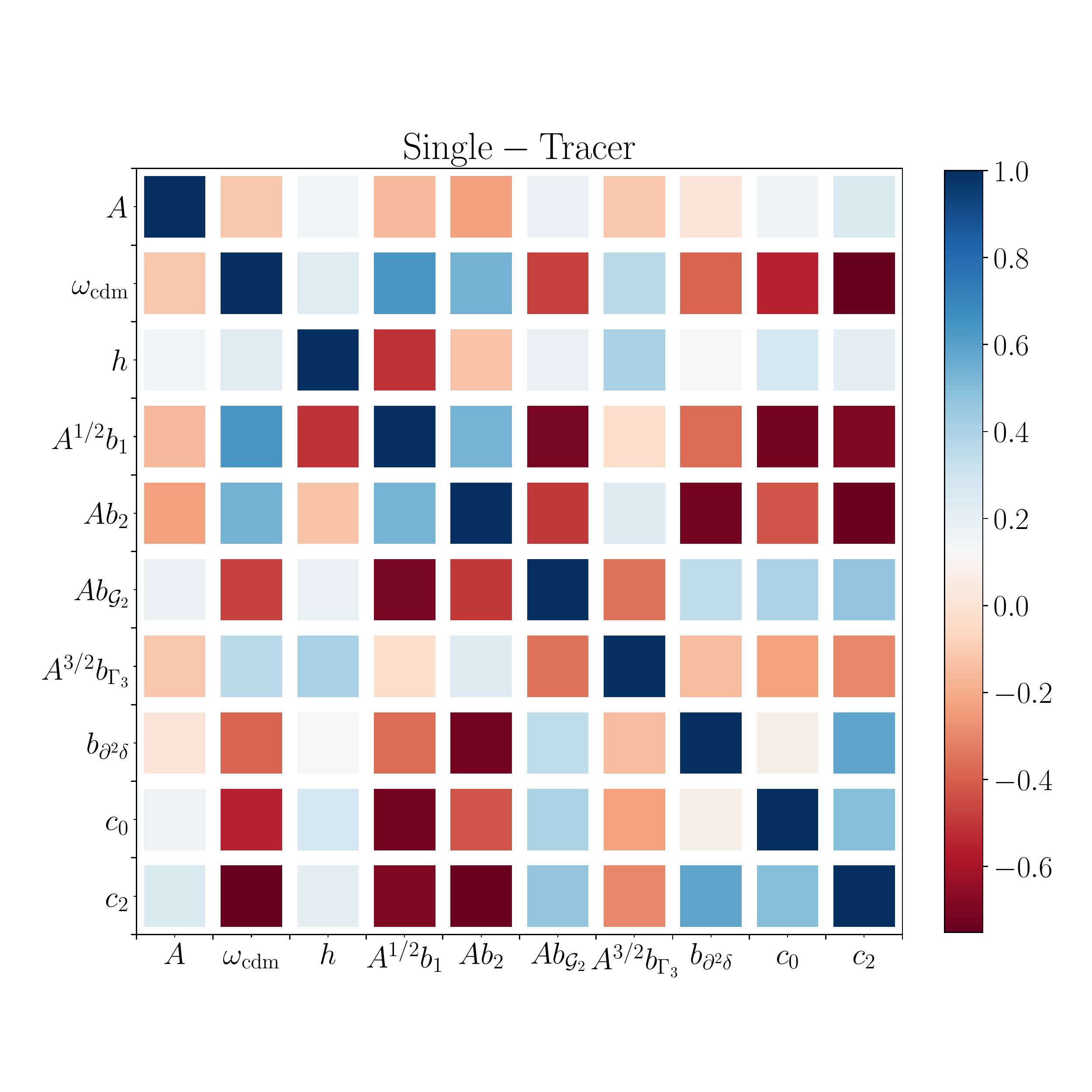}
    \includegraphics[width = 0.49 \textwidth]{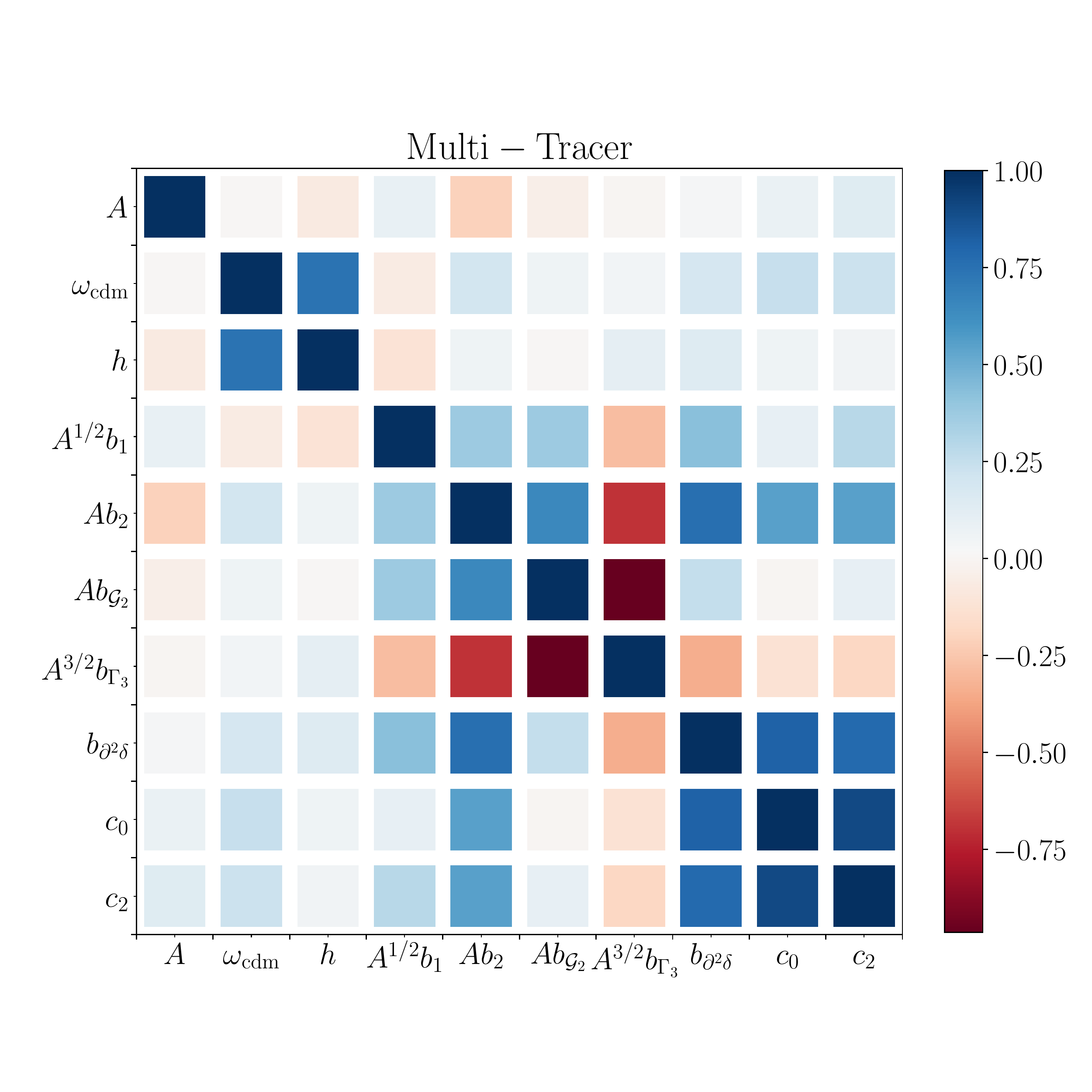}
    \caption{The correlation matrix for the $G_{0}$ prior and $k_{\rm max} = 0.14\, h\mathrm{Mpc}^{-1}$. The correlation matrix is computed using the MCMC posterior. In the left, for the single tracer parameters and in the right for the (effective) multi-tracer parameters.}
    \label{fig:correlation_matrix}
\end{figure}

When considering the MT scenario, the cross-spectrum provides a strong constraint on the combination $b^A_1b^B_1$, which then propagates as stronger constraints in $c_0$ in the self-spectra and also in the other free parameters. This can be better seen in the correlation matrices of the bias parameters for single and MTs, shown respectively by the left and right panel of Fig.~\ref{fig:correlation_matrix}. Notice that while the ST correlation matrix presents diagonal terms that are strongly correlated, the usage of MT has allowed us to break many degeneracies between the free parameters\footnote{Some degeneracies, however, do remain. We can see that $b_{\mathcal{G}_2}$ and $b_{\Gamma_3}$ are strongly correlated when using the power spectrum, which justifies the very-often exclusion of one of them from the analysis \cite{Wadekar:2020hax,Nishimichi:2020tvu}. Some degeneracies between bias parameters can not be broken by considering only the two-point functions \cite{saito2014understanding,Dizgah2020,Eggemeier2021}.}.

Finally, by splitting a halo catalogue into sub-populations we open the possibility to add/learn information about each sub-species (e.g. by using smaller mass bins or by taking into account assembly bias) \cite{Voivodic4}.
In Appendix~\ref{app: bias_coev} we present a scenario in which we add co-evolution relations to the priors.  
Co-evolution relations for the bias parameters are determined by the gravitational dynamics of the individual tracer species, and can in principle be used to improve the structure of the priors. 
However, we could not find strong evidence that the inclusion of co-evolution relations leads to improved constraints, at least at the level of power spectrum, both for single and multi-tracer.

%%%%%%%%%%%%%%%%. 

\section{Conclusions}
\label{sec:conclusion}

Extracting the maximal information out of large-scale structure is a key challenge that demands several different approaches. Extending the power spectrum analysis to higher n-point functions \cite{Sefusatti:2006pa, Gil-Marin:2016wya} allows us to probe the non-Gaussian component of the density field. Alternatively, attempts to consider non-linear maps of the density field (e.g. through the marked \cite{Philcox:2020fqx} and the log fields \cite{Rubira:2020inb}) have achieved remarkable results to improve cosmological constraints \cite{Massara:2020pli}. The fact that the matter density is probed indirectly via the bias expansion for tracers opens space for using combinations of  tracers to optimize the information content. 
In this work we have studied how multiple tracers can be used not only to beat down cosmic variance \cite{Seljak:2008xr}, but also to extract more information out of mildly non-linear scales via the EFTofLSS framework.

\begin{figure}[h!]
    \centering
    \includegraphics[width = 0.9\textwidth]{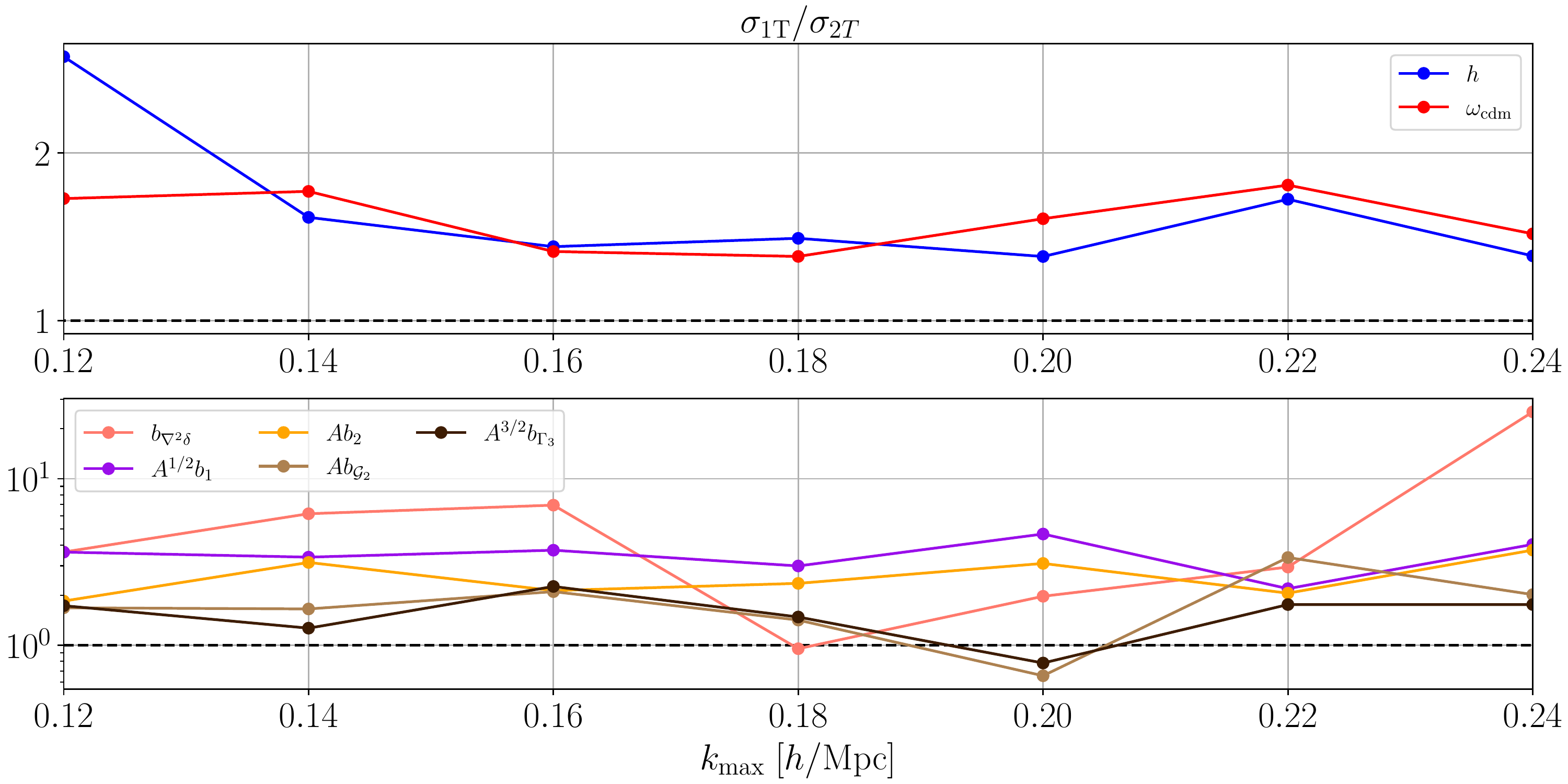}
    \caption{On top, the ratio between the $1\sigma$ confidence intervals of the single- and multi-tracer measurements for $\omega_{\rm cdm}$ and $h$ using the $G_{0}$ prior and for different values of $k_{\rm max}$. On bottom, the same but for the bias parameters.}
    \label{fig: overall_improvement}
\end{figure}

We divided a halo catalogue into two sub-catalogues by their masses, and used their self- and cross-spectra to constrain the cosmological and bias parameters. While splitting the tracer catalogue into sub-species brings in more information, this comes at the cost of a larger number of free parameters. However, the fact that the non-linear regime for those two sub-populations is non-degenerate (see Fig.~\ref{fig:bias_terms}) leads to overall information gain both for the \it effective \rm bias coefficients and for the cosmological parameters. We have shown in Appendix~\ref{app:galaxies} that when considering galaxies instead of halos this break of degeneracy is still valid and leads to substantial improvement in both $\Omega_{\rm cdm}$ and $h$. Another part of the improvement comes from the cross-spectrum, which was shown to break the degeneracy between $b_1$ and the other parameters (see Fig.~\ref{fig:correlation_matrix}). 
We summarize this effective gain in Fig.~\ref{fig: overall_improvement}, in which we display the ratio between the $1\sigma$ region for ST and MT for different parameters. For the cosmological parameters (top) we see that the error bars consistently shrink by a factor of $\sim$0.6 ($40\%$ improvement) for different values of $k_{\rm max}$, while for the bias parameters the gains can be even larger (bottom panel). When considering a simplified model without $b_{\Gamma_3}$ and $c_2$ the MT improvement drops to $40\%$. Besides, the results using MT have been shown to be less biased than the ST analysis.

We stress that the division into multiple tracers may open space for considering narrower priors onto the bias free parameters, motivated for instance by the study of other specific properties of their halos or their environments, such as their assembly bias. 
When considering a single tracer, many properties of those sub-populations might get averaged out, resulting in the loss of useful information.
In Appendix~\ref{app: bias_coev} we aimed at including additional information from the sub-species by considering the gravitational evolution of the tracers, but we could not find any substantial improvements in the resulting constraints.

In this paper we made evident that the usage of two tracers can substantially increase the information that can be extracted from the clustering of tracers, leading to improved constraints, fewer degeneracies, and smaller biases for some parameters. 
There are many distinct directions to explore after this work, which we leave for future studies, e.g. by considering higher-order n-point function, including redshift-space distortions, adding more information about the tracers through well-motivated theoretical priors, as well as splitting the halo catalogues into more tracers and testing the optimal division between them.

\acknowledgments
We acknowledge Florian Beutler for reading and commenting the manuscript. We also acknowledge the anonymous referee for constructive comments in the first version of the paper. TM would like to thank CNPq for financial support.
HR acknowledges the Deutsche Forschungsgemeinschaft under Germany's Excellence Strategy - EXC 2121 "Quantum Universe" - 390833306.
RV and RA thank the support of FAPESP through grants 2016/19647-2 and 2019/26492-3.

%%%%%%%%%%%%%%%%
\appendix
\section{Speeding up the MCMC by Taylor expanding} \label{app:taylor}

\paragraph{}Evaluating an MCMC-based analysis that involves cosmological parameter extraction is numerically expensive, since every step demands solving the Boltzmann system of equations for the linear theory, and calculating the one-loop integrals. Following \cite{Colas2019}, we performed a Taylor expansion over all perturbation theory diagrams  required to describe the tracer power spectrum up to one-loop with respect to the cosmological parameters $A_s$, $h$ and $\omega_{\rm cdm}$. By using the finite differences method up to second-order (stencil size $= 5$), we Taylor expanded up to the fourth order in each parameter. The reference value for the expansion as well as the relative step size for which we found the expansion to be efficient and stable are found in Table~\ref{Table: Taylor_setup}.

\begin{table}[hbt!]
\centering
\begin{tabular}{@{}ccc@{}}
\toprule
Cosmological parameter & Reference value       & Step size\\ \midrule
$A_s$                  & $2.14 \times 10^{-9}$ & 10\%  \\
$h$                    & $0.678$               & 5\%        \\
$\omega_{\rm cdm}$         & $0.119$               & 4\%       \\ \bottomrule
\end{tabular}
\caption{The setup used to Taylor expand the bias operators basis. The step size used in the finite difference derivation scheme is expressed as percentages of the reference value.}
\label{Table: Taylor_setup}
\end{table}

\begin{figure}[h]
    \centering
    \includegraphics[width = 0.8\textwidth]{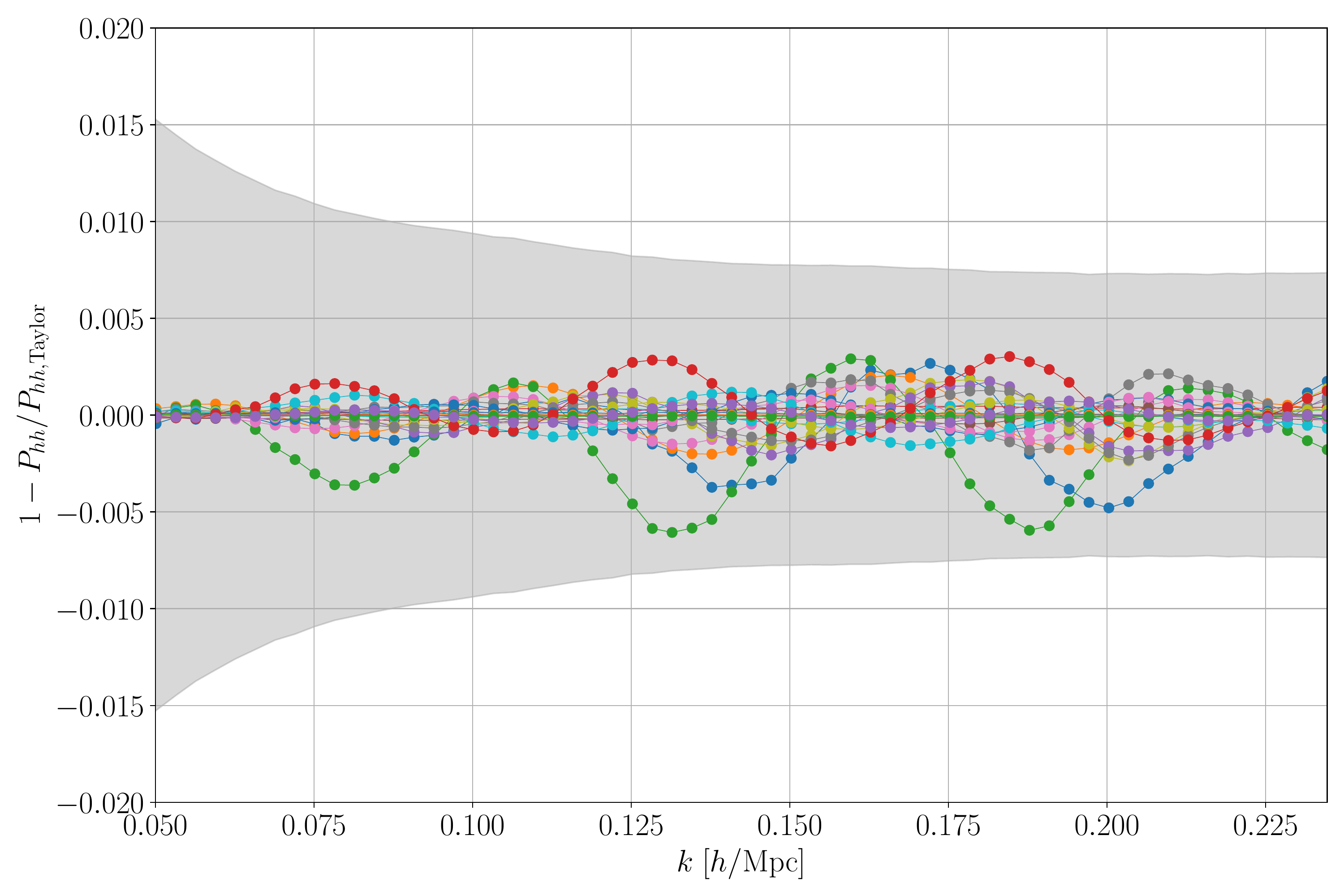}
    \caption{Relative error between the Taylor expanded and the exact power spectrum calculated by CLASS-PT. Each one of the 27 lines shows the spectrum computed in the range $A_{s,\rm fid}\; \pm\; 0.4 \times A_{s,\rm fid},\; h_{\rm fid}\; \pm\;  0.1\times h_{\rm fid},\; \omega_{\rm cdm, fid} \;\pm\;  0.13\times\omega_{\rm cdm, fid}$. The colored gray region is the error associated to the halo spectrum.}
    \label{fig: taylor_relative error}
\end{figure}
We found a time factor gain of $\approx 150$ for the Taylor expansion with respect to the usual \texttt{CLASS-PT} routine. We also emphasize the Taylor expansion was performed after the IR-Resummation.
 
 To test the accuracy of this Taylor expansion, we compare its output to the full evaluation of \texttt{CLASS-PT} in Fig.~\ref{fig: taylor_relative error}, where each line correspond to a different point in the parameter space. These points lie inside the intervals $A_{s,\rm fid}\; \pm\; 0.4 \times A_{s,\rm fid},\; h_{\rm fid}\; \pm\;  0.1\times h_{\rm fid},\; \omega_{\rm cdm, fid} \;\pm\;  0.13\times\omega_{\rm cdm, fid}$. This interval is wide enough for all set of cosmological parameters sampled in the MCMC analysis to lie well inside of it. We used this interval to define a three dimensional grid of three equally spaced points per side, what produces a total of 27 distinct set of points to be tested. The colored gray region is the error associated to the halo spectrum. Moreover, during a MCMC analysis, the Taylor expanded diagrams will be multiplied by the bias parameters that vary inside some specific range. It could happen that when some walker explores a region associated to large values of the bias parameters, the error of the Taylor expansion could be increased. To make sure we are safe from this problem, we performed the tests using values for the bias parameters that are higher than the typical ones we find in all analysis. Even in this situation, we concluded that our Taylor expanded diagrams are accurate enough to be used.

\section{Testing the bias co-evolution relations}
\label{app: bias_coev}

The splitting of a tracer into sub-populations opens space to add more information about the bias coefficients of the sub-species that could be blurred in the single tracer analysis. This information input can be translated in terms of narrower priors in the bias parameters. In this section we study the effect of considering an alternative prior to the bias than those pointed out in Table.~\ref{table:all_priors}. The idea is not exactly narrowing the priors but using information from the dynamics of the tracer to relate part of the bias coefficients (the so-called \it co-evolution relations\rm) and study how it affects the multi-tracer analysis.

\begin{table*}[t!h]
    \centering
    %%%%%%%%%%%%%%%%%%%%%%%%%%%%%%%%%%%%%%%%%%
    \begin{tabular}{|c||c|} \hline 
     &  Prior \; $G_{\rm co-ev}(\sigma)$\\ \hline \hline
    $b_1$  & Flat $[1.0, 2.2]$   \\ \hline
    $b_2$  & Gauss.(Eq.~\ref{eqn: bias_coev_b2},\,$\sigma$) \\ \hline 
    $b_{\mathcal{G}_2}$   & Gauss.(Eq.~\ref{eqn: bias_coev_bg2},\,$\sigma$) \\ \hline 
    $b_{\Gamma_3}$ &  Gauss.(Eq.~\ref{eqn: bias_coev_bGamma3},\,2$\sigma$) \\ \hline 
    $b_{\nabla^2\delta}$  &  Flat $[-50, 50]$\\ \hline
    $c_0$   & Flat $[-5.0, 5.0]$\\ \hline 
    $c_2$   & Flat $[-50, 50]$\\ \hline 
    \end{tabular}
    \caption{Priors over the bias and stochastic parameters for the $G_{\rm co-ev}(\sigma)$.}
    \label{table:co-ev_prior}
\end{table*}

The dynamics of structure formation imposes constraints on the bias expansion coefficients. 
We can understand those constraints by considering a multi-fluid component that satisfies the continuity equation\footnote{For simplicity we assume no velocity bias, and therefore Euler's equation is the same both for matter and for the tracer.} \cite{chan2012gravity, mirbabayi2015biased, saito2014understanding}:
given the initial conditions for the tracer at a formation time $\tau_{*}$, we can predict the evolution of the tracer field at later times. It implies that the dynamical component of the bias parameters is set by gravity. The initial values for the bias parameters (their value at formation time) are still however free parameters. 
A simplified version for the initial conditions assumes that, at the formation time, the tracer depends only on powers of $\delta$, an expansion which is often called  \it local-in-matter-density \rm (LIMD) \cite{Desjacques2016}. 
The subsequent nonlinear gravitational dynamics sources higher-derivatives operators $\mathcal{G}_2$ and $\Gamma_3$, which can be mapped into $b_1$ as \cite{Abidi2018,chan2012gravity}:
\begin{eqnarray}
    \label{eqn: bias_coev_bg2}
    b_{\mathcal{G}_2} &=& -\frac{2}{7}(b_1 - 1)\,,\\
    \label{eqn: bias_coev_bGamma3}
    b_{\Gamma_3} &=& \frac{23}{42}(b_1 - 1)\,.
\end{eqnarray}
Even though the validity of assuming vanishing $b_{\mathcal{G}_2}$ and $b_{\Gamma_3}$ at formation time is limited, as pointed out by \cite{Catelan:1997qw,Heavens:1998es,Smith:2006ne,chan2012gravity,Baldauf:2012hs,Barreira:2021ukk}, we here consider them as an approximation for the dynamical contribution.
For $b_2$, we will use a numerical fit from \cite{Lazeyras:2015lgp} obtained by using the separate Universe approach 
\begin{eqnarray}\label{eqn: bias_coev_b2}
    b_2 &=& 0.412-2.143\,b_1 + 0.929 \,(b_1)^2 + 0.008\, (b_1)^3\,.
\end{eqnarray}
Bearing in mind the limitations of the LIMD approximation, we do not rely on the reduced parameter hypersurface delimited by Eqs.~(\ref{eqn: bias_coev_bg2})-(\ref{eqn: bias_coev_b2}). 
Instead, analogously to what was done for $G_0$ in Sec.~\ref{sec:methodology}, we consider a Gaussian prior around Eqs.~(\ref{eqn: bias_coev_bg2})-(\ref{eqn: bias_coev_b2}) with variance $\sigma$. The priors considered in this appendix are described in Table~\ref{table:co-ev_prior} and we refer to them as $G_{\rm co-ev}$.

\begin{figure}[h]
    \centering
    \includegraphics[width = 0.9\textwidth]{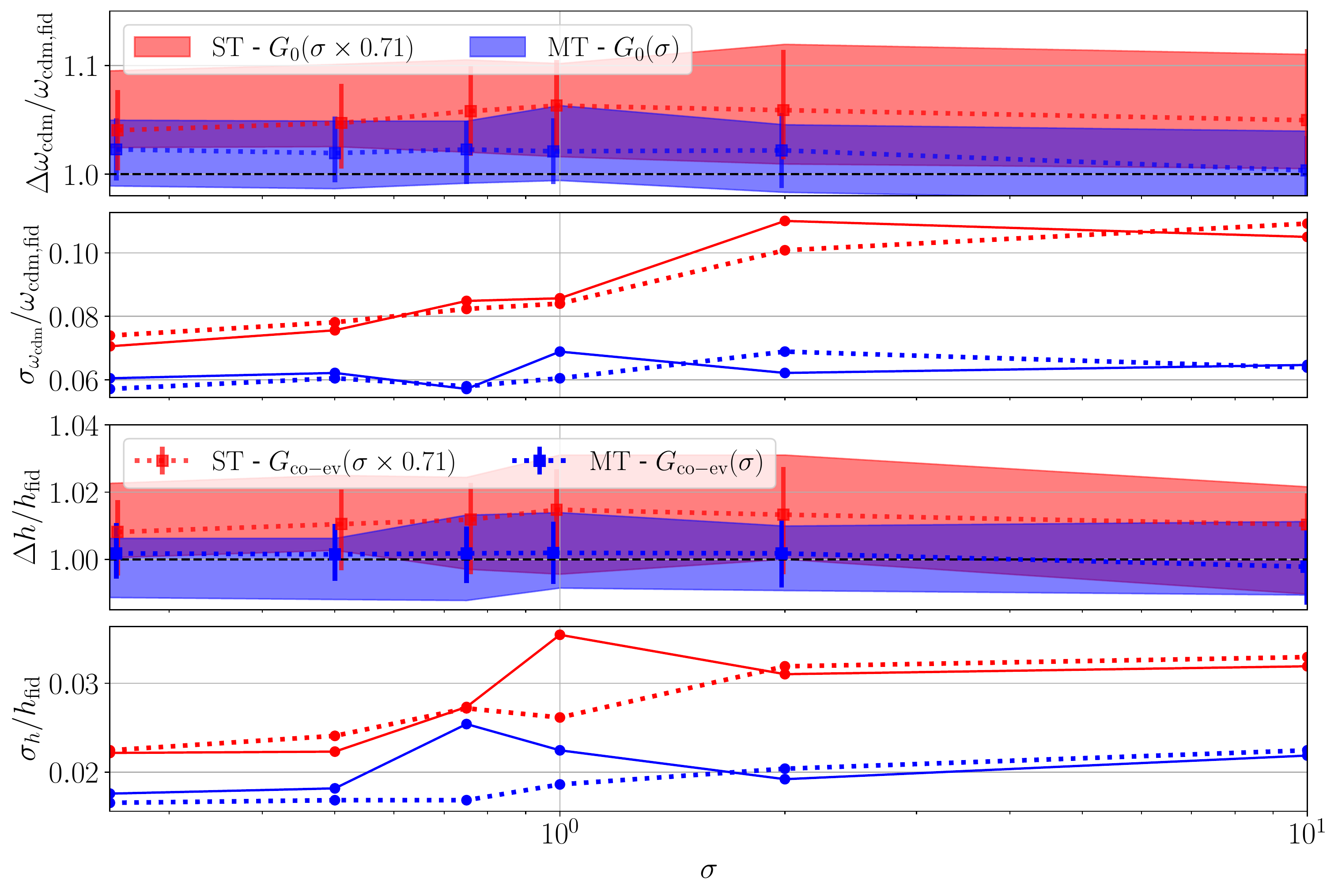}
    \caption{Cosmological parameters extracted by using the priors $G_0$ (shaded regions and solid lines) and $G_{\mathrm{co-ev}}$ (bars and dashed lines) as a function of $\sigma$ with $k_{\mathrm{max}} = 0.14 \, h\mathrm{Mpc}^{-1}$.}
    \label{fig: varying_sigma}
\end{figure}

\begin{figure}[h]
    \centering
    \includegraphics[width = 0.9\textwidth]{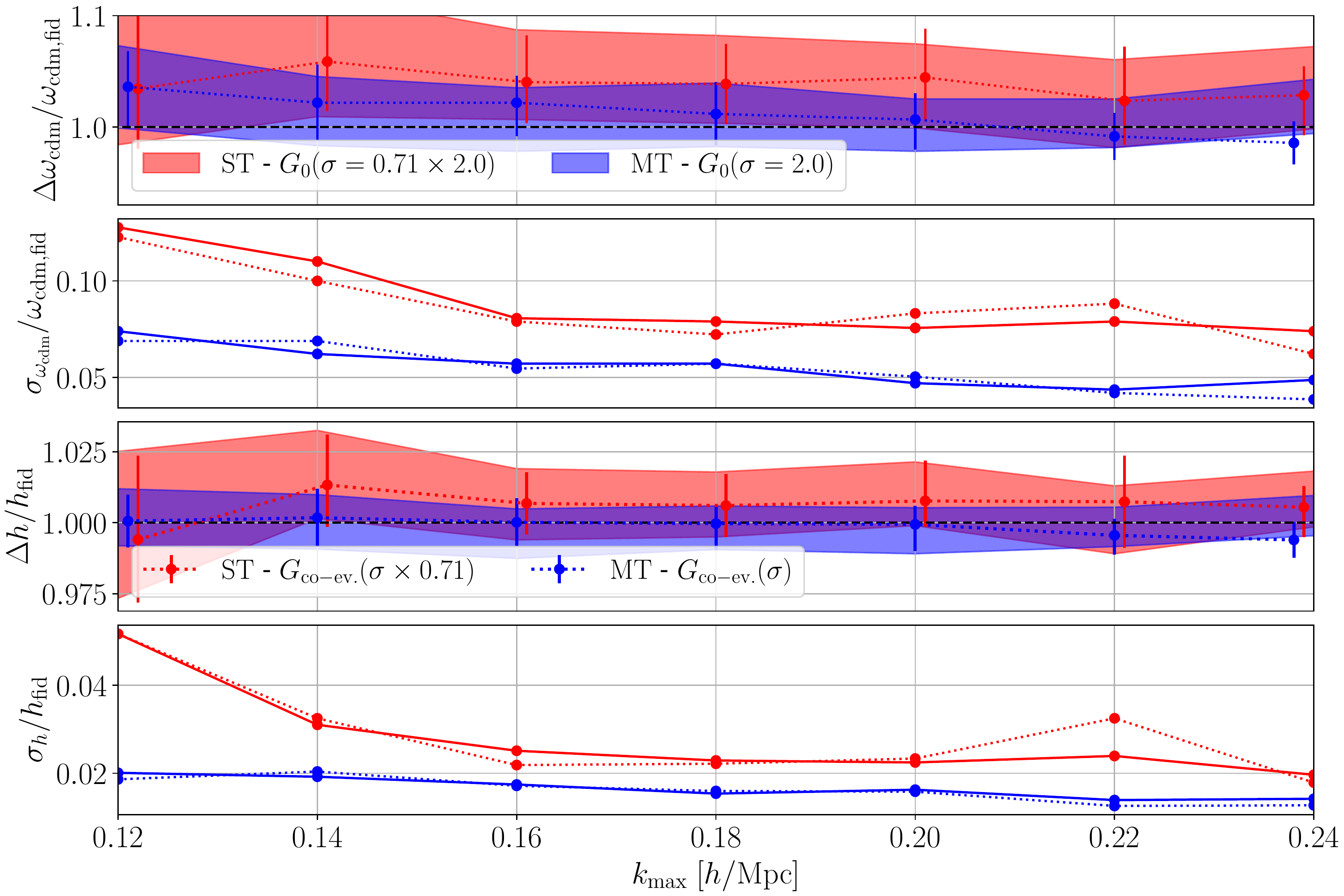}
    \caption{Comparison between $G_{\rm co-ev.}$ and $G_0$ with $\sigma=2$ as a function of $k_{\rm max}$.}
    \label{fig:coev_kmax}
\end{figure}

We show in Fig.~\ref{fig: varying_sigma} how the error bars for $h$, $\omega_{\rm cdm}$ and $b_1$ evolve as a function of the $\sigma$ used in the Gaussian priors of $b_2$, $b_{\mathcal{G}_2}$ and $b_{\Gamma_3}$, both for $G_{0}$ and $G_{\rm co-ev}$.  We have fixed $k_{\mathrm{max}} = 0.14 \,h\mathrm{Mpc}^{-1}$. We notice that, as expected, increasing $\sigma$ leads to a deterioration of constraints. For MT, however, this deterioration is less pronounced. Note also that MT provides a less-biased result for $\sigma \rightarrow 0$. We also note that the co-evolution relations provide a slightly better result for MT (which is not seen for ST), but this improvement is marginal. The Gaussian prior seems to have converged for $\sigma=2$, which we adopt as the fiducial value in the whole paper. 

We provide in Fig.~\ref{fig:coev_kmax} a comparison between $G_{\rm co-ev.}$ and $G_0$ with $\sigma=2$ for different values of $k_{\rm max}$. We can see that the effect of adding co-evolution is minimal both for MT and ST. We can see then that the co-evolution relations do \it not \rm affect the constraints coming from the power spectrum alone. Those relations however might still become relevant when considering higher-order n-point functions, trying to model the matter field directly as e.g., in the forward modelling \cite{Schmidt:2018bkr} or when extending the theory to higher $k_{\rm max}$.

\section{Analysis with galaxies}
\label{app:galaxies}

In this appendix we study with more attention the cross stochastic terms that were neglected in the analysis of the main text.

To consider a more realistic case, and a case where we expect a larger cross stochastic term, we populate the halos of the MultiDark simulation with galaxies. We fitted the halo occupation distribution using the IllustrisTNG300 simulation \cite{Pillepich:2017jle, 2017MNRAS.465.3291W, Nelson:2018uso}, similar to the way done in \cite{Voivodic4}. 

\begin{figure}[h]
    \centering
    \includegraphics[width=\textwidth]{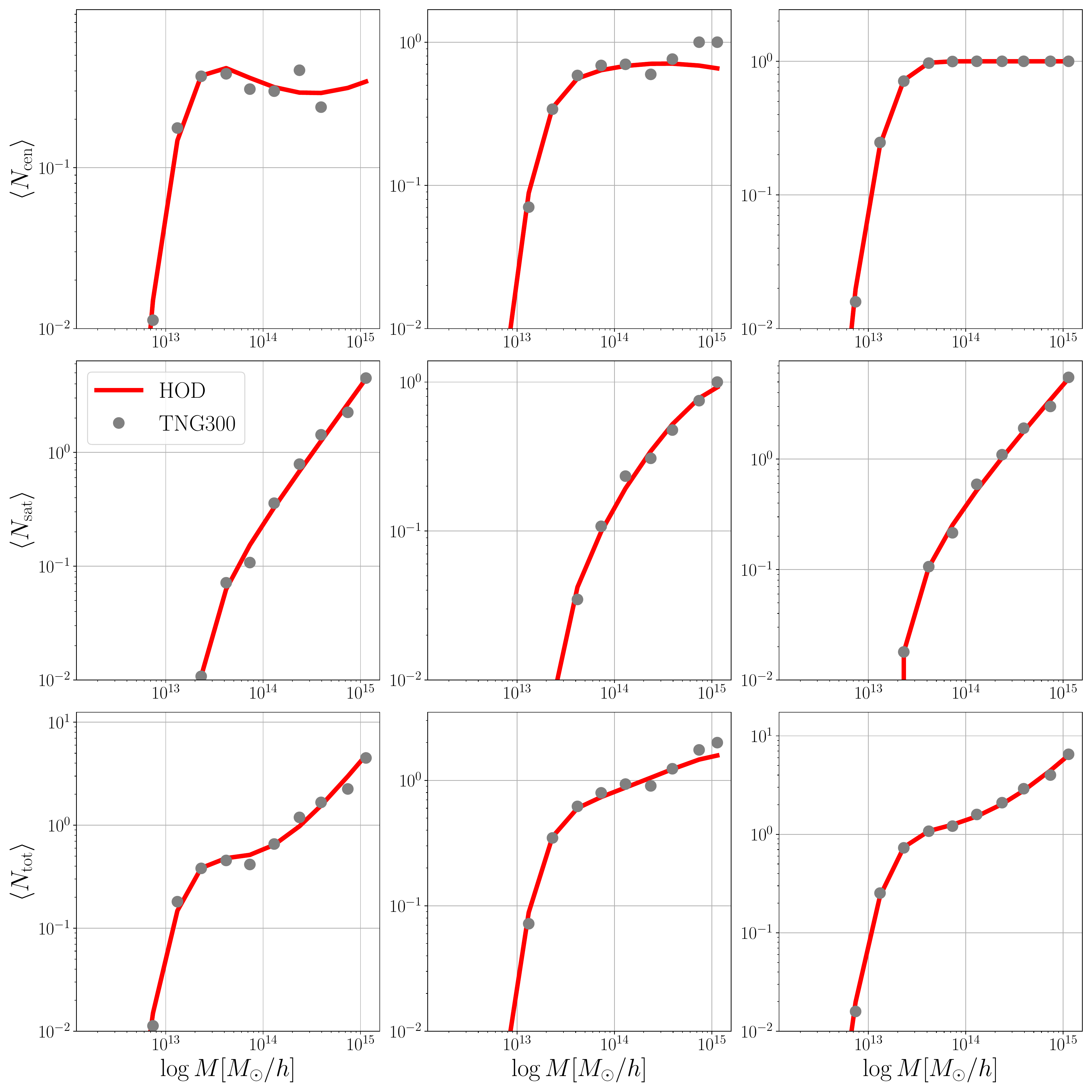}
    \caption{Comparison between the mean number of galaxies per bin of halo mass measured in Illustris-TNG300 (red dots) and the fitting function used to populate the dark matter halos (black lines). The first row shows the mean number for the central galaxies, the second row for the satellites and the third row the total mean number of galaxies. The first column presents the mean number of red galaxies (tracer A), the second column for the blue galaxies (tracer B) and the third one for the combination of the previous two (tracer A+B).}
    \label{fig:HOD_fit}
\end{figure}

We populated the halos using galaxies with $M_{\star} > 10^{11.2}$ $M_{\odot}/h$, to guarantee that halos less massive than the ones in our simulation have no galaxies. Then, we separate the galaxy sample into two populations, with $g-r < 0.75$ (blue galaxies) and $g-r > 0.75$ (red galaxies), where $g$ and $r$ are the magnitudes of each galaxy in the photometric bands $g$ and $r$, respectively, given by IllustrisTNG\footnote{Note that our separation in blue and red galaxies is not physical. We are in practice considering only red galaxies in this analysis and the cut in color is just a simple way to separate the galaxies in two different samples.}.

\begin{table*}[t!h]
    \centering
      \begin{tabular}{|c||c|c|c|} \hline 
               & $b_1^X$ &             $c^{XX}_0$   & $c^{XY}_0$ \\ \hline \hline
        Galaxy A     & 1.706 $\pm$ 0.018  & 0.28 $\pm$ 0.21 & 0.39 $\pm$ 0.19 \\ \hline
        Galaxy B     & 1.721 $\pm$ 0.012    & 0.51 $\pm$ 0.21 & 0.39 $\pm$ 0.19\\ \hline
    Galaxy A + B & 1.714 $\pm$ 0.017     & 0.77 $\pm$ 0.38 & - \\
    \hline
    \end{tabular}
    \caption{Same as Tab.~\ref{tab:fiducial_bias_shot} but with the bias and stochastic parameters for the galaxies.}
    \label{tab:fiducial_bias_shot_galaxies} 
\end{table*}

In Fig.~\ref{fig:HOD_fit} we show the mean number of galaxies as a function of the host halo mass for the two populations of galaxies we consider here (blue and red galaxies). We also show the separation between central (first row), satellites (second row), and the total sample of galaxies (third row). The red dots show the measurements from IllustrisTNG300 and the black lines the fitting function used to describe the mean halo occupation distribution of these galaxies. 

This HOD was fitted using the expression from \cite{2007ApJ...667..760Z}, for the occupancy of total galaxies, and a second order polynomial for the relative number of red and blue galaxies. The galaxies were distributed using the NFW profile \cite{Navarro:1995iw} with the exclusion term proposed in \cite{2020JCAP...10..033V} with $\sigma = 0.5$. We can see a good agreement between the fitting function and the simulation data.

\begin{figure}[h]
    \centering
    \includegraphics[width = \textwidth]{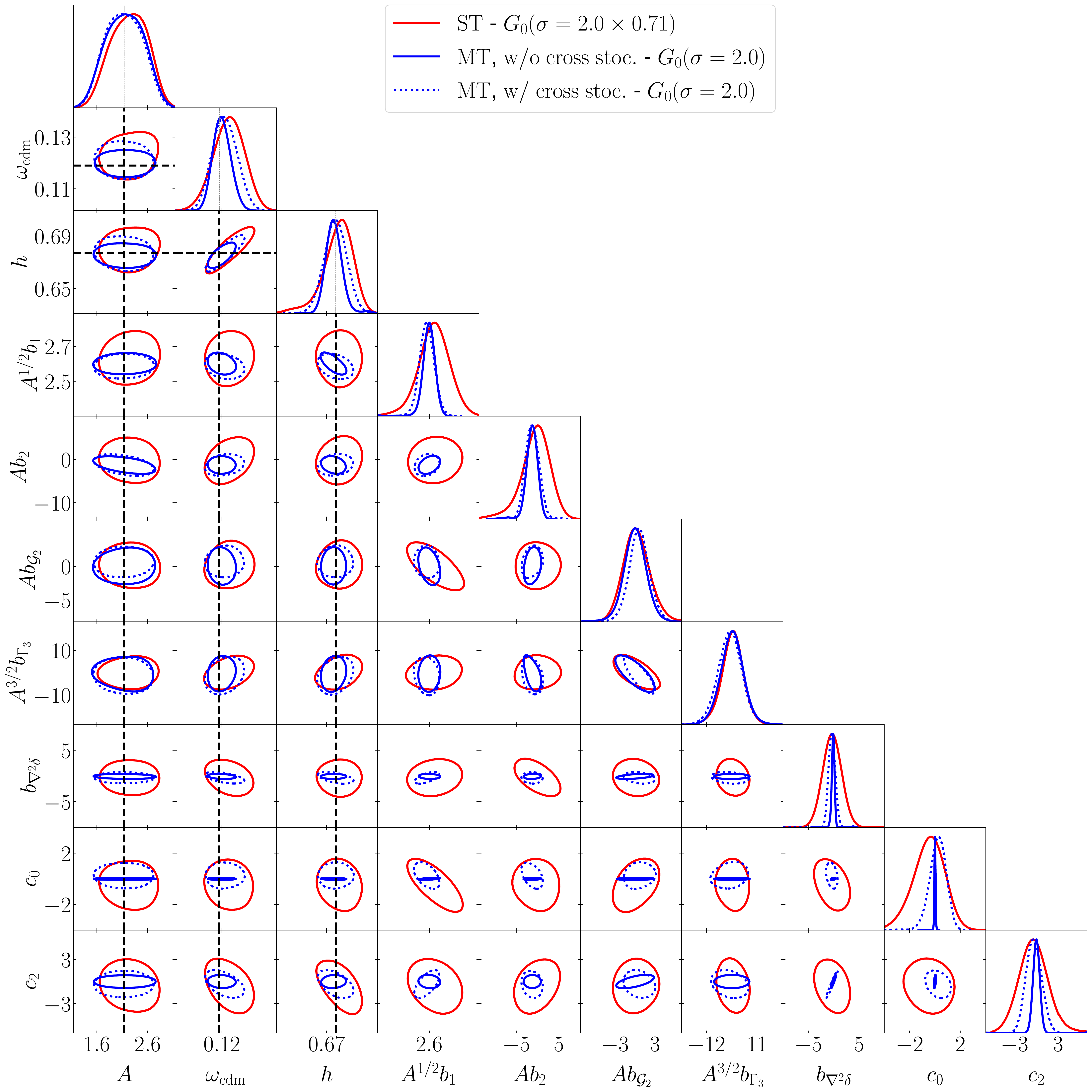}
    \caption{Same as Fig.~\ref{fig:mcmc_result_G0} but for the two galaxy catalogues generated by populating the halos with a HOD.}
    \label{fig:mcmc_result_G0_galaxie}
\end{figure}

By considering two populations of galaxies separated by color and not two populations of halos separated by their masses, we are mixing the masses of the host halos. In that way, we expect to find a non-vanishing stochastic term in the cross spectrum \cite{2021arXiv211200012K}.

In Tab.~\ref{tab:fiducial_bias_shot_galaxies} we show the linear bias and first-order stochastic parameters for the three galaxy samples (blue, red and blue+red) in the $k\rightarrow0$ limit (see Eqs.~(\ref{eqn: low_k_auto}) and (\ref{eqn: low_k_cross})). For the case of galaxies, in contrast to the case of halos (Tab.~\ref{tab:fiducial_bias_shot}), we see that the stochastic term of the cross spectrum is different from zero in more than two sigmas. Therefore, this term is non-vanishing and should not be neglected.

Fig.~\ref{fig:mcmc_result_G0_galaxie} shows the same results as in Fig.~\ref{fig:mcmc_result_G0} but for our galaxy catalogue. We show the results considering only one tracer (red lines), the results considering two tracers and fixing the cross stochastic term in zero (blue solid line), and considering the cross stochastic term in the analysis (dotted blue line).
Similar to the case of halos, when we fix the cross stochastic terms the constraints are improved, especially for $b_{1}$, $b_{\nabla ^{2} \delta}$ and the stochastic terms. However, even fixing these non-vanishing parameters, we see that the constraints are not biased with respect to the case considering the full parameter space. 

Therefore, our results for the halos are not dependent of the consideration of the stochastic parameters in the analysis, once they are consistent with zero (see Tab.~\ref{tab:fiducial_bias_shot}) and, even when they are not, they do not bias the final constraints (see Fig.~\ref{fig:mcmc_result_G0_galaxie}). Using galaxies under the consideration of a non-vanishing cross stochastic term still improves the constraints in the cosmological parameters ($21\%$ in $\omega_{\rm cdm}$ and $22\%$ in $h$). It resembles the fact that most part of the gain in the full-shape analysis of MT comes from the break of degeneracy in the (mild) non-linear part of the spectra, as pointed out in Fig.~\ref{fig:bias_terms}. When we divide galaxies according to their color we checked that their spectra are still non-degenerate. We leave a more detailed analysis for a forthcoming paper \cite{Mergulhao:2022}.

\bibliographystyle{JHEP}
\bibliography{main}

\providecommand{\href}[2]{#2}\begingroup\raggedright\begin{thebibliography}{10}

\bibitem{Abbetal}
{The Dark Energy Survey Collaboration}, \emph{{The Dark Energy Survey}},
  {\emph{ArXiv Astrophysics e-prints} (Oct., 2005) },
  [\href{http://arxiv.org/abs/astro-ph/0510346}{{\tt astro-ph/0510346}}].

\bibitem{Amendola:2012ys}
{\scshape Euclid Theory Working Group} collaboration, L.~Amendola et~al.,
  \emph{{Cosmology and fundamental physics with the Euclid satellite}},
  \href{http://dx.doi.org/10.12942/lrr-2013-6}{\emph{Living Rev. Rel.} {\bf 16}
  (2013) 6}, [\href{http://arxiv.org/abs/1206.1225}{{\tt 1206.1225}}].

\bibitem{Benitez:2014ibt}
{\scshape J-PAS} collaboration, N.~Benitez et~al., \emph{{J-PAS: The
  Javalambre-Physics of the Accelerated Universe Astrophysical Survey}},
  \href{http://arxiv.org/abs/1403.5237}{{\tt 1403.5237}}.

\bibitem{Ivezic:2008fe}
{\scshape LSST} collaboration, Z.~Ivezic, J.~A. Tyson, R.~Allsman, J.~Andrew
  and R.~Angel, \emph{{LSST: from Science Drivers to Reference Design and
  Anticipated Data Products}},  \href{http://arxiv.org/abs/0805.2366}{{\tt
  0805.2366}}.

\bibitem{Beutler:2019ojk}
F.~Beutler, M.~Biagetti, D.~Green, A.~Slosar and B.~Wallisch, \emph{{Primordial
  Features from Linear to Nonlinear Scales}},
  \href{http://dx.doi.org/10.1103/PhysRevResearch.1.033209}{\emph{Phys. Rev.
  Res.} {\bf 1} (2019) 033209}, [\href{http://arxiv.org/abs/1906.08758}{{\tt
  1906.08758}}].

\bibitem{Baumann:2010tm}
D.~Baumann, A.~Nicolis, L.~Senatore and M.~Zaldarriaga, \emph{{Cosmological
  Non-Linearities as an Effective Fluid}},
  \href{http://dx.doi.org/10.1088/1475-7516/2012/07/051}{\emph{JCAP} {\bf 07}
  (2012) 051}, [\href{http://arxiv.org/abs/1004.2488}{{\tt 1004.2488}}].

\bibitem{Carrasco:2012cv}
J.~J.~M. Carrasco, M.~P. Hertzberg and L.~Senatore, \emph{{The Effective Field
  Theory of Cosmological Large Scale Structures}},
  \href{http://dx.doi.org/10.1007/JHEP09(2012)082}{\emph{JHEP} {\bf 09} (2012)
  082}, [\href{http://arxiv.org/abs/1206.2926}{{\tt 1206.2926}}].

\bibitem{Carrasco:2013mua}
J.~J.~M. Carrasco, S.~Foreman, D.~Green and L.~Senatore, \emph{{The Effective
  Field Theory of Large Scale Structures at Two Loops}},
  \href{http://dx.doi.org/10.1088/1475-7516/2014/07/057}{\emph{JCAP} {\bf 07}
  (2014) 057}, [\href{http://arxiv.org/abs/1310.0464}{{\tt 1310.0464}}].

\bibitem{Konstandin:2019bay}
T.~Konstandin, R.~A. Porto and H.~Rubira, \emph{{The Effective Field Theory of
  Large Scale Structure at Three Loops}},
  \href{http://dx.doi.org/10.1088/1475-7516/2019/11/027}{\emph{JCAP} {\bf 11}
  (2019) 027}, [\href{http://arxiv.org/abs/1906.00997}{{\tt 1906.00997}}].

\bibitem{Angulo:2015eqa}
R.~Angulo, M.~Fasiello, L.~Senatore and Z.~Vlah, \emph{{On the Statistics of
  Biased Tracers in the Effective Field Theory of Large Scale Structures}},
  \href{http://dx.doi.org/10.1088/1475-7516/2015/9/029}{\emph{JCAP} {\bf 09}
  (2015) 029}, [\href{http://arxiv.org/abs/1503.08826}{{\tt 1503.08826}}].

\bibitem{DAmico:2019fhj}
G.~D'Amico, J.~Gleyzes, N.~Kokron, K.~Markovic, L.~Senatore, P.~Zhang et~al.,
  \emph{{The Cosmological Analysis of the SDSS/BOSS data from the Effective
  Field Theory of Large-Scale Structure}},
  \href{http://dx.doi.org/10.1088/1475-7516/2020/05/005}{\emph{JCAP} {\bf 05}
  (2020) 005}, [\href{http://arxiv.org/abs/1909.05271}{{\tt 1909.05271}}].

\bibitem{Ivanov:2019hqk}
M.~M. Ivanov, M.~Simonovi\'c and M.~Zaldarriaga, \emph{{Cosmological Parameters
  and Neutrino Masses from the Final Planck and Full-Shape BOSS Data}},
  \href{http://dx.doi.org/10.1103/PhysRevD.101.083504}{\emph{Phys. Rev. D} {\bf
  101} (2020) 083504}, [\href{http://arxiv.org/abs/1912.08208}{{\tt
  1912.08208}}].

\bibitem{Colas:2019ret}
T.~Colas, G.~D'amico, L.~Senatore, P.~Zhang and F.~Beutler, \emph{{Efficient
  Cosmological Analysis of the SDSS/BOSS data from the Effective Field Theory
  of Large-Scale Structure}},
  \href{http://dx.doi.org/10.1088/1475-7516/2020/06/001}{\emph{JCAP} {\bf 06}
  (2020) 001}, [\href{http://arxiv.org/abs/1909.07951}{{\tt 1909.07951}}].

\bibitem{Philcox:2020vvt}
O.~H. Philcox, M.~M. Ivanov, M.~Simonovi\'c and M.~Zaldarriaga,
  \emph{{Combining Full-Shape and BAO Analyses of Galaxy Power Spectra: A 1.6\%
  CMB-independent constraint on H0}},
  \href{http://dx.doi.org/10.1088/1475-7516/2020/05/032}{\emph{JCAP} {\bf 05}
  (2020) 032}, [\href{http://arxiv.org/abs/2002.04035}{{\tt 2002.04035}}].

\bibitem{Nishimichi:2020tvu}
T.~Nishimichi, G.~D'Amico, M.~M. Ivanov, L.~Senatore, M.~Simonovi\'c, M.~Takada
  et~al., \emph{{Blinded challenge for precision cosmology with large-scale
  structure: results from effective field theory for the redshift-space galaxy
  power spectrum}},
  \href{http://dx.doi.org/10.1103/PhysRevD.102.123541}{\emph{Phys. Rev. D} {\bf
  102} (2020) 123541}, [\href{http://arxiv.org/abs/2003.08277}{{\tt
  2003.08277}}].

\bibitem{Ivanov:2020ril}
M.~M. Ivanov, E.~McDonough, J.~C. Hill, M.~Simonovi\'c, M.~W. Toomey,
  S.~Alexander et~al., \emph{{Constraining Early Dark Energy with Large-Scale
  Structure}}, \href{http://dx.doi.org/10.1103/PhysRevD.102.103502}{\emph{Phys.
  Rev. D} {\bf 102} (2020) 103502},
  [\href{http://arxiv.org/abs/2006.11235}{{\tt 2006.11235}}].

\bibitem{Lague:2021frh}
A.~Lagu\"e, J.~R. Bond, R.~Hlo\v{z}ek, K.~K. Rogers, D.~J.~E. Marsh and
  D.~Grin, \emph{{Constraining Ultralight Axions with Galaxy Surveys}},
  \href{http://arxiv.org/abs/2104.07802}{{\tt 2104.07802}}.

\bibitem{Assassi2014}
V.~Assassi, D.~Baumann, D.~Green and M.~Zaldarriaga, \emph{{Renormalized Halo
  Bias}}, \href{http://dx.doi.org/10.1088/1475-7516/2014/08/056}{\emph{JCAP}
  {\bf 08} (2014) 056}, [\href{http://arxiv.org/abs/1402.5916}{{\tt
  1402.5916}}].

\bibitem{Desjacques2016}
V.~Desjacques, D.~Jeong and F.~Schmidt, \emph{{Large-Scale Galaxy Bias}},
  \href{http://dx.doi.org/10.1016/j.physrep.2017.12.002}{\emph{Phys. Rep.} {\bf
  733} (nov, 2016) 1--193}, [\href{http://arxiv.org/abs/1611.09787}{{\tt
  1611.09787}}].

\bibitem{Tegmark:1998wm}
M.~Tegmark and P.~J.~E. Peebles, \emph{{The Time evolution of bias}},
  \href{http://dx.doi.org/10.1086/311426}{\emph{Astrophys. J. Lett.} {\bf 500}
  (1998) L79}, [\href{http://arxiv.org/abs/astro-ph/9804067}{{\tt
  astro-ph/9804067}}].

\bibitem{Modi:2016dah}
C.~Modi, E.~Castorina and U.~Seljak, \emph{{Halo bias in Lagrangian Space:
  Estimators and theoretical predictions}},
  \href{http://dx.doi.org/10.1093/mnras/stx2148}{\emph{Mon. Not. Roy. Astron.
  Soc.} {\bf 472} (2017) 3959--3970},
  [\href{http://arxiv.org/abs/1612.01621}{{\tt 1612.01621}}].

\bibitem{McDonald:2009dh}
P.~McDonald and A.~Roy, \emph{{Clustering of dark matter tracers: generalizing
  bias for the coming era of precision LSS}},
  \href{http://dx.doi.org/10.1088/1475-7516/2009/08/020}{\emph{JCAP} {\bf 08}
  (2009) 020}, [\href{http://arxiv.org/abs/0902.0991}{{\tt 0902.0991}}].

\bibitem{mirbabayi2015biased}
M.~Mirbabayi, F.~Schmidt and M.~Zaldarriaga, \emph{{Biased Tracers and Time
  Evolution}},
  \href{http://dx.doi.org/10.1088/1475-7516/2015/07/030}{\emph{JCAP} {\bf 07}
  (2015) 030}, [\href{http://arxiv.org/abs/1412.5169}{{\tt 1412.5169}}].

\bibitem{Lazeyras:2017hxw}
T.~Lazeyras and F.~Schmidt, \emph{{Beyond LIMD bias: a measurement of the
  complete set of third-order halo bias parameters}},
  \href{http://dx.doi.org/10.1088/1475-7516/2018/09/008}{\emph{JCAP} {\bf 09}
  (2018) 008}, [\href{http://arxiv.org/abs/1712.07531}{{\tt 1712.07531}}].

\bibitem{Schmittfull:2018yuk}
M.~Schmittfull, M.~Simonovi\'c, V.~Assassi and M.~Zaldarriaga, \emph{{Modeling
  Biased Tracers at the Field Level}},
  \href{http://dx.doi.org/10.1103/PhysRevD.100.043514}{\emph{Phys. Rev. D} {\bf
  100} (2019) 043514}, [\href{http://arxiv.org/abs/1811.10640}{{\tt
  1811.10640}}].

\bibitem{Baldauf:2013hka}
T.~Baldauf, U.~Seljak, R.~E. Smith, N.~Hamaus and V.~Desjacques, \emph{{Halo
  stochasticity from exclusion and nonlinear clustering}},
  \href{http://dx.doi.org/10.1103/PhysRevD.88.083507}{\emph{Phys. Rev. D} {\bf
  88} (2013) 083507}, [\href{http://arxiv.org/abs/1305.2917}{{\tt 1305.2917}}].

\bibitem{Seljak:2008xr}
U.~Seljak, \emph{{Extracting primordial non-gaussianity without cosmic
  variance}},
  \href{http://dx.doi.org/10.1103/PhysRevLett.102.021302}{\emph{Phys. Rev.
  Lett.} {\bf 102} (2009) 021302}, [\href{http://arxiv.org/abs/0807.1770}{{\tt
  0807.1770}}].

\bibitem{McDonald:2008sh}
P.~McDonald and U.~Seljak, \emph{{How to measure redshift-space distortions
  without sample variance}},
  \href{http://dx.doi.org/10.1088/1475-7516/2009/10/007}{\emph{JCAP} {\bf 10}
  (2009) 007}, [\href{http://arxiv.org/abs/0810.0323}{{\tt 0810.0323}}].

\bibitem{Abramo2013}
L.~R. Abramo and K.~E. Leonard, \emph{{Why multi-tracer surveys beat cosmic
  variance}}, \href{http://dx.doi.org/10.1093/mnras/stt465}{\emph{Mon. Not.
  Roy. Astron. Soc.} {\bf 432} (2013) 318},
  [\href{http://arxiv.org/abs/1302.5444}{{\tt 1302.5444}}].

\bibitem{zhao2020completed}
Y.~Wang et~al., \emph{{The clustering of the SDSS-IV extended Baryon
  Oscillation Spectroscopic Survey DR16 luminous red galaxy and emission line
  galaxy samples: cosmic distance and structure growth measurements using
  multiple tracers in configuration space}},
  \href{http://dx.doi.org/10.1093/mnras/staa2593}{\emph{Mon. Not. Roy. Astron.
  Soc.} {\bf 498} (2020) 3470--3483},
  [\href{http://arxiv.org/abs/2007.09010}{{\tt 2007.09010}}].

\bibitem{Montero-Dorta:2019kyb}
A.~D. Montero-Dorta, L.~R. Abramo, B.~R. Granett, S.~de~la Torre and L.~Guzzo,
  \emph{{The Multi-Tracer Optimal Estimator applied to VIPERS}},
  \href{http://dx.doi.org/10.1093/mnras/staa405}{\emph{Mon. Not. Roy. Astron.
  Soc.} {\bf 493} (2020) 5257--5272},
  [\href{http://arxiv.org/abs/1909.00010}{{\tt 1909.00010}}].

\bibitem{favole2019cosmological}
G.~Favole, D.~Sapone and J.~S. Lafaurie, \emph{{Cosmological constraints from
  galaxy multi-tracers in the nearby Universe}},
  \href{http://arxiv.org/abs/1912.06155}{{\tt 1912.06155}}.

\bibitem{klypin2016multidark}
A.~Klypin, G.~Yepes, S.~Gottlober, F.~Prada and S.~Hess, \emph{{MultiDark
  simulations: the story of dark matter halo concentrations and density
  profiles}}, \href{http://dx.doi.org/10.1093/mnras/stw248}{\emph{Mon. Not.
  Roy. Astron. Soc.} {\bf 457} (2016) 4340--4359},
  [\href{http://arxiv.org/abs/1411.4001}{{\tt 1411.4001}}].

\bibitem{2008JCAP...08..031S}
A.~{Slosar}, C.~{Hirata}, U.~{Seljak}, S.~{Ho} and N.~{Padmanabhan},
  \emph{{Constraints on local primordial non-Gaussianity from large scale
  structure}},
  \href{http://dx.doi.org/10.1088/1475-7516/2008/08/031}{\emph{\jcap} {\bf
  2008} (Aug., 2008) 031}, [\href{http://arxiv.org/abs/0805.3580}{{\tt
  0805.3580}}].

\bibitem{2010CQGra..27l4011D}
V.~{Desjacques} and U.~{Seljak}, \emph{{Primordial non-Gaussianity from the
  large-scale structure}},
  \href{http://dx.doi.org/10.1088/0264-9381/27/12/124011}{\emph{Classical and
  Quantum Gravity} {\bf 27} (June, 2010) 124011},
  [\href{http://arxiv.org/abs/1003.5020}{{\tt 1003.5020}}].

\bibitem{Hamaus:2010im}
N.~Hamaus, U.~Seljak, V.~Desjacques, R.~E. Smith and T.~Baldauf,
  \emph{{Minimizing the Stochasticity of Halos in Large-Scale Structure
  Surveys}}, \href{http://dx.doi.org/10.1103/PhysRevD.82.043515}{\emph{Phys.
  Rev. D} {\bf 82} (2010) 043515}, [\href{http://arxiv.org/abs/1004.5377}{{\tt
  1004.5377}}].

\bibitem{Hamaus:2011dq}
N.~Hamaus, U.~Seljak and V.~Desjacques, \emph{{Optimal Constraints on Local
  Primordial Non-Gaussianity from the Two-Point Statistics of Large-Scale
  Structure}}, \href{http://dx.doi.org/10.1103/PhysRevD.84.083509}{\emph{Phys.
  Rev. D} {\bf 84} (2011) 083509}, [\href{http://arxiv.org/abs/1104.2321}{{\tt
  1104.2321}}].

\bibitem{Hamaus:2012ap}
N.~Hamaus, U.~Seljak and V.~Desjacques, \emph{{Optimal Weighting in Galaxy
  Surveys: Application to Redshift-Space Distortions}},
  \href{http://dx.doi.org/10.1103/PhysRevD.86.103513}{\emph{Phys. Rev. D} {\bf
  86} (2012) 103513}, [\href{http://arxiv.org/abs/1207.1102}{{\tt 1207.1102}}].

\bibitem{2010MNRAS.407..772G}
H.~{Gil-Mar{\'\i}n}, C.~{Wagner}, L.~{Verde}, R.~{Jimenez} and A.~F. {Heavens},
  \emph{{Reducing sample variance: halo biasing, non-linearity and
  stochasticity}},
  \href{http://dx.doi.org/10.1111/j.1365-2966.2010.16958.x}{\emph{\mnras} {\bf
  407} (Sept., 2010) 772--790}, [\href{http://arxiv.org/abs/1003.3238}{{\tt
  1003.3238}}].

\bibitem{2011MNRAS.416.3009B}
G.~M. {Bernstein} and Y.-C. {Cai}, \emph{{Cosmology without cosmic variance}},
  \href{http://dx.doi.org/10.1111/j.1365-2966.2011.19249.x}{\emph{\mnras} {\bf
  416} (Oct., 2011) 3009--3016}, [\href{http://arxiv.org/abs/1104.3862}{{\tt
  1104.3862}}].

\bibitem{ginzburg2020shot}
D.~Ginzburg and V.~Desjacques, \emph{{Shot noise in multitracer constraints on
  fNL and relativistic projections: Power spectrum}},
  \href{http://dx.doi.org/10.1093/mnras/staa1154}{\emph{Mon. Not. Roy. Astron.
  Soc.} {\bf 495} (2020) 932--942},
  [\href{http://arxiv.org/abs/1911.11701}{{\tt 1911.11701}}].

\bibitem{2017PhRvD..96l3535A}
L.~R. {Abramo} and D.~{Bertacca}, \emph{{Disentangling the effects of Doppler
  velocity and primordial non-Gaussianity in galaxy power spectra}},
  \href{http://dx.doi.org/10.1103/PhysRevD.96.123535}{\emph{\prd} {\bf 96}
  (Dec., 2017) 123535}, [\href{http://arxiv.org/abs/1706.01834}{{\tt
  1706.01834}}].

\bibitem{2013MNRAS.436.3089B}
C.~{Blake}, I.~K. {Baldry}, J.~{Bland-Hawthorn}, L.~{Christodoulou},
  M.~{Colless}, C.~{Conselice} et~al., \emph{{Galaxy And Mass Assembly (GAMA):
  improved cosmic growth measurements using multiple tracers of large-scale
  structure}}, \href{http://dx.doi.org/10.1093/mnras/stt1791}{\emph{\mnras}
  {\bf 436} (Dec., 2013) 3089--3105},
  [\href{http://arxiv.org/abs/1309.5556}{{\tt 1309.5556}}].

\bibitem{wang2020clustering}
Y.~Wang et~al., \emph{{The clustering of the SDSS-IV extended Baryon
  Oscillation Spectroscopic Survey DR16 luminous red galaxy and emission line
  galaxy samples: cosmic distance and structure growth measurements using
  multiple tracers in configuration space}},
  \href{http://dx.doi.org/10.1093/mnras/staa2593}{\emph{Mon. Not. Roy. Astron.
  Soc.} {\bf 498} (2020) 3470--3483},
  [\href{http://arxiv.org/abs/2007.09010}{{\tt 2007.09010}}].

\bibitem{2014MNRAS.442.2511F}
L.~D. {Ferramacho}, M.~G. {Santos}, M.~J. {Jarvis} and S.~{Camera},
  \emph{{Radio galaxy populations and the multitracer technique: pushing the
  limits on primordial non-Gaussianity}},
  \href{http://dx.doi.org/10.1093/mnras/stu1015}{\emph{\mnras} {\bf 442} (Aug.,
  2014) 2511--2518}, [\href{http://arxiv.org/abs/1402.2290}{{\tt 1402.2290}}].

\bibitem{2015ApJ...803...21B}
P.~{Bull}, P.~G. {Ferreira}, P.~{Patel} and M.~G. {Santos}, \emph{{Late-time
  Cosmology with 21 cm Intensity Mapping Experiments}},
  \href{http://dx.doi.org/10.1088/0004-637X/803/1/21}{\emph{\apj} {\bf 803}
  (Apr., 2015) 21}, [\href{http://arxiv.org/abs/1405.1452}{{\tt 1405.1452}}].

\bibitem{tanidis2020developing}
K.~Tanidis and S.~Camera, \emph{{Developing a unified pipeline for large-scale
  structure data analysis with angular power spectra \textendash{} III.
  Implementing the multitracer technique to constrain neutrino masses}},
  \href{http://dx.doi.org/10.1093/mnras/staa3536}{\emph{Mon. Not. Roy. Astron.
  Soc.} {\bf 502} (2021) 2952--2960},
  [\href{http://arxiv.org/abs/2009.05584}{{\tt 2009.05584}}].

\bibitem{wang2020brief}
Y.~Wang and G.-B. Zhao, \emph{{A brief review on cosmological analysis of
  galaxy surveys with multiple tracers}},
  \href{http://arxiv.org/abs/2009.03862}{{\tt 2009.03862}}.

\bibitem{viljoen2020constraining}
J.-A. Viljoen, J.~Fonseca and R.~Maartens, \emph{{Constraining the growth rate
  by combining multiple future surveys}},
  \href{http://dx.doi.org/10.1088/1475-7516/2020/09/054}{\emph{JCAP} {\bf 09}
  (2020) 054}, [\href{http://arxiv.org/abs/2007.04656}{{\tt 2007.04656}}].

\bibitem{liu2021coupling}
R.~H. Liu and P.~C. Breysse, \emph{{Coupling parsec and gigaparsec scales:
  Primordial non-Gaussianity with multitracer intensity mapping}},
  \href{http://dx.doi.org/10.1103/PhysRevD.103.063520}{\emph{Phys. Rev. D} {\bf
  103} (2021) 063520}, [\href{http://arxiv.org/abs/2002.10483}{{\tt
  2002.10483}}].

\bibitem{gomes2020non}
Z.~Gomes, S.~Camera, M.~J. Jarvis, C.~Hale and J.~Fonseca,
  \emph{{Non-Gaussianity constraints using future radio continuum surveys and
  the multitracer technique}},
  \href{http://dx.doi.org/10.1093/mnras/stz3581}{\emph{Mon. Not. Roy. Astron.
  Soc.} {\bf 492} (2020) 1513--1522},
  [\href{http://arxiv.org/abs/1912.08362}{{\tt 1912.08362}}].

\bibitem{2016arXiv161100036D}
{DESI Collaboration}, A.~{Aghamousa}, J.~{Aguilar}, S.~{Ahlen}, S.~{Alam},
  L.~E. {Allen} et~al., \emph{{The DESI Experiment Part I: Science,Targeting,
  and Survey Design}}, {\emph{arXiv e-prints} (Oct., 2016) arXiv:1611.00036},
  [\href{http://arxiv.org/abs/1611.00036}{{\tt 1611.00036}}].

\bibitem{2014arXiv1403.5237B}
N.~{Benitez}, R.~{Dupke}, M.~{Moles}, L.~{Sodre}, J.~{Cenarro},
  A.~{Marin-Franch} et~al., \emph{{J-PAS: The Javalambre-Physics of the
  Accelerated Universe Astrophysical Survey}}, {\emph{arXiv e-prints} (Mar.,
  2014) arXiv:1403.5237}, [\href{http://arxiv.org/abs/1403.5237}{{\tt
  1403.5237}}].

\bibitem{2020MNRAS.493.3616A}
M.~{Aparicio Resco}, A.~L. {Maroto}, J.~S. {Alcaniz}, L.~R. {Abramo},
  C.~{Hern{\'a}ndez-Monteagudo}, N.~{Ben{\'\i}tez} et~al., \emph{{J-PAS:
  forecasts on dark energy and modified gravity theories}},
  \href{http://dx.doi.org/10.1093/mnras/staa367}{\emph{\mnras} {\bf 493} (Apr.,
  2020) 3616--3631}, [\href{http://arxiv.org/abs/1910.02694}{{\tt
  1910.02694}}].

\bibitem{2016MNRAS.455.3871A}
L.~R. {Abramo}, L.~F. {Secco} and A.~{Loureiro}, \emph{{Fourier analysis of
  multitracer cosmological surveys}},
  \href{http://dx.doi.org/10.1093/mnras/stv2588}{\emph{\mnras} {\bf 455} (Feb.,
  2016) 3871--3889}, [\href{http://arxiv.org/abs/1505.04106}{{\tt
  1505.04106}}].

\bibitem{Abramo2012}
L.~R. {Abramo}, \emph{{The full Fisher matrix for galaxy surveys}},
  \href{http://dx.doi.org/10.1111/j.1365-2966.2011.20166.x}{\emph{\mnras} {\bf
  420} (Mar., 2012) 2042--2057}, [\href{http://arxiv.org/abs/1108.5449}{{\tt
  1108.5449}}].

\bibitem{chudaykin2020nonlinear}
A.~Chudaykin, M.~M. Ivanov, O.~H.~E. Philcox and M.~Simonovi\'c,
  \emph{{Nonlinear perturbation theory extension of the Boltzmann code CLASS}},
  \href{http://dx.doi.org/10.1103/PhysRevD.102.063533}{\emph{Phys. Rev. D} {\bf
  102} (2020) 063533}, [\href{http://arxiv.org/abs/2004.10607}{{\tt
  2004.10607}}].

\bibitem{Bernardeau2001}
F.~Bernardeau, S.~Colombi, E.~Gaztanaga and R.~Scoccimarro, \emph{{Large-Scale
  Structure of the Universe and Cosmological Perturbation Theory}},
  \href{http://dx.doi.org/10.1016/S0370-1573(02)00135-7}{\emph{Phys. Rep.} {\bf
  367} (dec, 2001) 1--248}, [\href{http://arxiv.org/abs/0112551}{{\tt
  0112551}}].

\bibitem{Senatore:2014via}
L.~Senatore and M.~Zaldarriaga, \emph{{The IR-resummed Effective Field Theory
  of Large Scale Structures}},
  \href{http://dx.doi.org/10.1088/1475-7516/2015/02/013}{\emph{JCAP} {\bf 02}
  (2015) 013}, [\href{http://arxiv.org/abs/1404.5954}{{\tt 1404.5954}}].

\bibitem{Eggemeier2020}
A.~Eggemeier, R.~Scoccimarro, M.~Crocce, A.~Pezzotta and A.~G. S\'anchez,
  \emph{{Testing one-loop galaxy bias: Power spectrum}},
  \href{http://dx.doi.org/10.1103/PhysRevD.102.103530}{\emph{Phys. Rev. D} {\bf
  102} (2020) 103530}, [\href{http://arxiv.org/abs/2006.09729}{{\tt
  2006.09729}}].

\bibitem{Eggemeier2021}
A.~Eggemeier, R.~Scoccimarro, R.~E. Smith, M.~Crocce, A.~Pezzotta and A.~G.
  S\'anchez, \emph{{Testing one-loop galaxy bias: Joint analysis of power
  spectrum and bispectrum}},
  \href{http://dx.doi.org/10.1103/PhysRevD.103.123550}{\emph{Phys. Rev. D} {\bf
  103} (2021) 123550}, [\href{http://arxiv.org/abs/2102.06902}{{\tt
  2102.06902}}].

\bibitem{Ivanov2019}
M.~M. Ivanov, M.~Simonovi\'c and M.~Zaldarriaga, \emph{{Cosmological Parameters
  from the BOSS Galaxy Power Spectrum}},
  \href{http://dx.doi.org/10.1088/1475-7516/2020/05/042}{\emph{JCAP} {\bf 05}
  (2020) 042}, [\href{http://arxiv.org/abs/1909.05277}{{\tt 1909.05277}}].

\bibitem{Wadekar:2020hax}
D.~Wadekar, M.~M. Ivanov and R.~Scoccimarro, \emph{{Cosmological constraints
  from BOSS with analytic covariance matrices}},
  \href{http://dx.doi.org/10.1103/PhysRevD.102.123521}{\emph{Phys. Rev. D} {\bf
  102} (2020) 123521}, [\href{http://arxiv.org/abs/2009.00622}{{\tt
  2009.00622}}].

\bibitem{Blot_2019}
L.~Blot, M.~Crocce, E.~Sefusatti, M.~Lippich, A.~G. Sánchez, M.~Colavincenzo
  et~al., \emph{Comparing approximate methods for mock catalogues and
  covariance matrices ii: power spectrum multipoles},
  \href{http://dx.doi.org/10.1093/mnras/stz507}{\emph{Monthly Notices of the
  Royal Astronomical Society} {\bf 485} (Feb, 2019) 2806–2824}.

\bibitem{Wadekar:2019rdu}
D.~Wadekar and R.~Scoccimarro, \emph{{Galaxy power spectrum multipoles
  covariance in perturbation theory}},
  \href{http://dx.doi.org/10.1103/PhysRevD.102.123517}{\emph{Phys. Rev. D} {\bf
  102} (2020) 123517}, [\href{http://arxiv.org/abs/1910.02914}{{\tt
  1910.02914}}].

\bibitem{2013PASP..125..306F}
D.~{Foreman-Mackey}, D.~W. {Hogg}, D.~{Lang} and J.~{Goodman}, \emph{{emcee:
  The MCMC Hammer}}, \href{http://dx.doi.org/10.1086/670067}{\emph{\pasp} {\bf
  125} (Mar., 2013) 306}, [\href{http://arxiv.org/abs/1202.3665}{{\tt
  1202.3665}}].

\bibitem{Lewis:2019xzd}
A.~Lewis, \emph{{GetDist: a Python package for analysing Monte Carlo samples}},
   \href{http://arxiv.org/abs/1910.13970}{{\tt 1910.13970}}.

\bibitem{Colas2019}
T.~Colas, G.~D'Amico, L.~Senatore, P.~Zhang and F.~Beutler, \emph{{Efficient
  Cosmological Analysis of the SDSS/BOSS data from the Effective Field Theory
  of Large-Scale Structure}},  \href{http://arxiv.org/abs/1909.07951}{{\tt
  1909.07951}}.

\bibitem{Baldauf:2016sjb}
T.~Baldauf, M.~Mirbabayi, M.~Simonovi\'c and M.~Zaldarriaga, \emph{{LSS
  constraints with controlled theoretical uncertainties}},
  \href{http://arxiv.org/abs/1602.00674}{{\tt 1602.00674}}.

\bibitem{Cooray:2002dia}
A.~Cooray and R.~K. Sheth, \emph{{Halo Models of Large Scale Structure}},
  \href{http://dx.doi.org/10.1016/S0370-1573(02)00276-4}{\emph{Phys. Rept.}
  {\bf 372} (2002) 1--129}, [\href{http://arxiv.org/abs/astro-ph/0206508}{{\tt
  astro-ph/0206508}}].

\bibitem{2021arXiv211200012K}
N.~{Kokron}, J.~{DeRose}, S.-F. {Chen}, M.~{White} and R.~H. {Wechsler},
  \emph{{Priors on red galaxy stochasticity from hybrid effective field
  theory}}, {\emph{arXiv e-prints} (Nov., 2021) arXiv:2112.00012},
  [\href{http://arxiv.org/abs/2112.00012}{{\tt 2112.00012}}].

\bibitem{saito2014understanding}
S.~Saito, T.~Baldauf, Z.~Vlah, U.~Seljak, T.~Okumura and P.~McDonald,
  \emph{{Understanding higher-order nonlocal halo bias at large scales by
  combining the power spectrum with the bispectrum}},
  \href{http://dx.doi.org/10.1103/PhysRevD.90.123522}{\emph{Phys. Rev. D} {\bf
  90} (2014) 123522}, [\href{http://arxiv.org/abs/1405.1447}{{\tt 1405.1447}}].

\bibitem{Dizgah2020}
A.~Moradinezhad~Dizgah, M.~Biagetti, E.~Sefusatti, V.~Desjacques and
  J.~Nore\~na, \emph{{Primordial Non-Gaussianity from Biased Tracers:
  Likelihood Analysis of Real-Space Power Spectrum and Bispectrum}},
  \href{http://dx.doi.org/10.1088/1475-7516/2021/05/015}{\emph{JCAP} {\bf 05}
  (2021) 015}, [\href{http://arxiv.org/abs/2010.14523}{{\tt 2010.14523}}].

\bibitem{Voivodic4}
R.~{Voivodic} and A.~{Barreira}, \emph{{Responses of Halo Occupation
  Distributions: a new ingredient in the halo model \& the impact on galaxy
  bias}}, {\emph{arXiv e-prints} (Dec., 2020) arXiv:2012.04637},
  [\href{http://arxiv.org/abs/2012.04637}{{\tt 2012.04637}}].

\bibitem{Sefusatti:2006pa}
E.~Sefusatti, M.~Crocce, S.~Pueblas and R.~Scoccimarro, \emph{{Cosmology and
  the Bispectrum}},
  \href{http://dx.doi.org/10.1103/PhysRevD.74.023522}{\emph{Phys. Rev. D} {\bf
  74} (2006) 023522}, [\href{http://arxiv.org/abs/astro-ph/0604505}{{\tt
  astro-ph/0604505}}].

\bibitem{Gil-Marin:2016wya}
H.~Gil-Mar\'\i{}n, W.~J. Percival, L.~Verde, J.~R. Brownstein, C.-H. Chuang,
  F.-S. Kitaura et~al., \emph{{The clustering of galaxies in the SDSS-III
  Baryon Oscillation Spectroscopic Survey: RSD measurement from the power
  spectrum and bispectrum of the DR12 BOSS galaxies}},
  \href{http://dx.doi.org/10.1093/mnras/stw2679}{\emph{Mon. Not. Roy. Astron.
  Soc.} {\bf 465} (2017) 1757--1788},
  [\href{http://arxiv.org/abs/1606.00439}{{\tt 1606.00439}}].

\bibitem{Philcox:2020fqx}
O.~H.~E. Philcox, E.~Massara and D.~N. Spergel, \emph{{What does the marked
  power spectrum measure? Insights from perturbation theory}},
  \href{http://dx.doi.org/10.1103/PhysRevD.102.043516}{\emph{Phys. Rev. D} {\bf
  102} (2020) 043516}, [\href{http://arxiv.org/abs/2006.10055}{{\tt
  2006.10055}}].

\bibitem{Rubira:2020inb}
H.~Rubira and R.~Voivodic, \emph{{The Effective Field Theory and Perturbative
  Analysis for Log-Density Fields}},
  \href{http://dx.doi.org/10.1088/1475-7516/2021/03/070}{\emph{JCAP} {\bf 03}
  (2021) 070}, [\href{http://arxiv.org/abs/2011.12280}{{\tt 2011.12280}}].

\bibitem{Massara:2020pli}
E.~Massara, F.~Villaescusa-Navarro, S.~Ho, N.~Dalal and D.~N. Spergel,
  \emph{{Using the Marked Power Spectrum to Detect the Signature of Neutrinos
  in Large-Scale Structure}},
  \href{http://dx.doi.org/10.1103/PhysRevLett.126.011301}{\emph{Phys. Rev.
  Lett.} {\bf 126} (2021) 011301}, [\href{http://arxiv.org/abs/2001.11024}{{\tt
  2001.11024}}].

\bibitem{chan2012gravity}
K.~C. Chan, R.~Scoccimarro and R.~K. Sheth, \emph{{Gravity and Large-Scale
  Non-local Bias}},
  \href{http://dx.doi.org/10.1103/PhysRevD.85.083509}{\emph{Phys. Rev. D} {\bf
  85} (2012) 083509}, [\href{http://arxiv.org/abs/1201.3614}{{\tt 1201.3614}}].

\bibitem{Abidi2018}
M.~M. Abidi and T.~Baldauf, \emph{{Cubic Halo Bias in Eulerian and Lagrangian
  Space}}, \href{http://dx.doi.org/10.1088/1475-7516/2018/07/029}{\emph{JCAP}
  {\bf 07} (2018) 029}, [\href{http://arxiv.org/abs/1802.07622}{{\tt
  1802.07622}}].

\bibitem{Catelan:1997qw}
P.~Catelan, F.~Lucchin, S.~Matarrese and C.~Porciani, \emph{{The bias field of
  dark matter halos}},
  \href{http://dx.doi.org/10.1046/j.1365-8711.1998.01455.x}{\emph{Mon. Not.
  Roy. Astron. Soc.} {\bf 297} (1998) 692--712},
  [\href{http://arxiv.org/abs/astro-ph/9708067}{{\tt astro-ph/9708067}}].

\bibitem{Heavens:1998es}
A.~F. Heavens, S.~Matarrese and L.~Verde, \emph{{The Nonlinear redshift-space
  power spectrum of galaxies}},
  \href{http://dx.doi.org/10.1046/j.1365-8711.1998.02052.x}{\emph{Mon. Not.
  Roy. Astron. Soc.} {\bf 301} (1998) 797--808},
  [\href{http://arxiv.org/abs/astro-ph/9808016}{{\tt astro-ph/9808016}}].

\bibitem{Smith:2006ne}
R.~E. Smith, R.~Scoccimarro and R.~K. Sheth, \emph{{The Scale Dependence of
  Halo and Galaxy Bias: Effects in Real Space}},
  \href{http://dx.doi.org/10.1103/PhysRevD.75.063512}{\emph{Phys. Rev. D} {\bf
  75} (2007) 063512}, [\href{http://arxiv.org/abs/astro-ph/0609547}{{\tt
  astro-ph/0609547}}].

\bibitem{Baldauf:2012hs}
T.~Baldauf, U.~Seljak, V.~Desjacques and P.~McDonald, \emph{{Evidence for
  Quadratic Tidal Tensor Bias from the Halo Bispectrum}},
  \href{http://dx.doi.org/10.1103/PhysRevD.86.083540}{\emph{Phys. Rev. D} {\bf
  86} (2012) 083540}, [\href{http://arxiv.org/abs/1201.4827}{{\tt 1201.4827}}].

\bibitem{Barreira:2021ukk}
A.~Barreira, T.~Lazeyras and F.~Schmidt, \emph{{Galaxy bias from forward
  models: linear and second-order bias of IllustrisTNG galaxies}},
  \href{http://arxiv.org/abs/2105.02876}{{\tt 2105.02876}}.

\bibitem{Lazeyras:2015lgp}
T.~Lazeyras, C.~Wagner, T.~Baldauf and F.~Schmidt, \emph{{Precision measurement
  of the local bias of dark matter halos}},
  \href{http://dx.doi.org/10.1088/1475-7516/2016/02/018}{\emph{JCAP} {\bf 02}
  (2016) 018}, [\href{http://arxiv.org/abs/1511.01096}{{\tt 1511.01096}}].

\bibitem{Schmidt:2018bkr}
F.~Schmidt, F.~Elsner, J.~Jasche, N.~M. Nguyen and G.~Lavaux, \emph{{A rigorous
  EFT-based forward model for large-scale structure}},
  \href{http://dx.doi.org/10.1088/1475-7516/2019/01/042}{\emph{JCAP} {\bf 01}
  (2019) 042}, [\href{http://arxiv.org/abs/1808.02002}{{\tt 1808.02002}}].

\bibitem{Pillepich:2017jle}
A.~Pillepich et~al., \emph{{Simulating Galaxy Formation with the IllustrisTNG
  Model}}, \href{http://dx.doi.org/10.1093/mnras/stx2656}{\emph{Mon. Not. Roy.
  Astron. Soc.} {\bf 473} (2018) 4077--4106},
  [\href{http://arxiv.org/abs/1703.02970}{{\tt 1703.02970}}].

\bibitem{2017MNRAS.465.3291W}
R.~{Weinberger}, V.~{Springel}, L.~{Hernquist}, A.~{Pillepich}, F.~{Marinacci},
  R.~{Pakmor} et~al., \emph{{Simulating galaxy formation with black hole driven
  thermal and kinetic feedback}},
  \href{http://dx.doi.org/10.1093/mnras/stw2944}{\emph{\mnras} {\bf 465} (Mar,
  2017) 3291--3308}, [\href{http://arxiv.org/abs/1607.03486}{{\tt
  1607.03486}}].

\bibitem{Nelson:2018uso}
D.~Nelson et~al., \emph{{The IllustrisTNG Simulations: Public Data Release}},
  {\emph{arXiv:1812.05609} (2018) },
  [\href{http://arxiv.org/abs/1812.05609}{{\tt 1812.05609}}].

\bibitem{2007ApJ...667..760Z}
Z.~{Zheng}, A.~L. {Coil} and I.~{Zehavi}, \emph{{Galaxy Evolution from Halo
  Occupation Distribution Modeling of DEEP2 and SDSS Galaxy Clustering}},
  \href{http://dx.doi.org/10.1086/521074}{\emph{\apj} {\bf 667} (Oct., 2007)
  760--779}, [\href{http://arxiv.org/abs/astro-ph/0703457}{{\tt
  astro-ph/0703457}}].

\bibitem{Navarro:1995iw}
J.~F. Navarro, C.~S. Frenk and S.~D.~M. White, \emph{{The Structure of cold
  dark matter halos}}, \href{http://dx.doi.org/10.1086/177173}{\emph{Astrophys.
  J.} {\bf 462} (1996) 563--575},
  [\href{http://arxiv.org/abs/astro-ph/9508025}{{\tt astro-ph/9508025}}].

\bibitem{2020JCAP...10..033V}
R.~{Voivodic}, H.~{Rubira} and M.~{Lima}, \emph{{The Halo Void (Dust) Model of
  large scale structure}},
  \href{http://dx.doi.org/10.1088/1475-7516/2020/10/033}{\emph{\jcap} {\bf
  2020} (Oct., 2020) 033}, [\href{http://arxiv.org/abs/2003.06411}{{\tt
  2003.06411}}].

\bibitem{Mergulhao:2022}
T.~Mergulh\~ao, H.~Rubira, R.~Voivodic and L.~R. Abramo, \emph{{The Effective
  Field Theory of Large-Scale Structure and Multi-tracer: connection to
  observations}},  \href{http://arxiv.org/abs/In preparation}{{\tt In
  preparation}}.

\end{thebibliography}\endgroup

\end{document}